\documentclass[12pt]{article}
\usepackage[T1]{fontenc}
\usepackage{baskervald}

\usepackage[font=small, labelfont=bf]{caption}
\usepackage{graphicx}
\usepackage{hyperref}
\usepackage{amsmath}
\usepackage{amssymb}
\usepackage{amsthm}
\usepackage{longtable}
\usepackage{verbatim}
\usepackage{rotating}
\usepackage{afterpage}

\usepackage{pifont}
\newcommand{\cmark}{\ding{51}}%
\newcommand{\xmark}{\ding{55}}%

\usepackage{cite}

\usepackage[usenames,dvipsnames]{xcolor}
\usepackage{tikz}
\usetikzlibrary{trees}
\usetikzlibrary{decorations.pathmorphing}
\usetikzlibrary{decorations.markings}
\usetikzlibrary{arrows}

\theoremstyle{definition}

\newcommand{\be}{\begin{equation}}
\newcommand{\ee}{\end{equation}}
\newcommand{\beq}{\begin{eqnarray}}
\newcommand{\eeq}{\end{eqnarray}}
\def\simlt{\stackrel{<}{{}_\sim}}
\def\simgt{\stackrel{>}{{}_\sim}}
\newcommand{\btheta}{\frac{\theta}{2\pi}}

\usepackage{titlesec}

\title{Weak Gravity Strongly Constrains\\ Large-Field Axion Inflation}

\author{Ben Heidenreich, Matthew Reece, and Tom Rudelius\\
{\small \color{gray} \texttt{bjheiden, mreece, rudelius~(@physics.harvard.edu)}}\\
Department of Physics, Harvard University, Cambridge, MA, 02138}

\begin{document}
\maketitle

\begin{abstract}
Models of large-field inflation based on axion-like fields with shift symmetries can be simple and natural, and make a promising prediction of detectable primordial gravitational waves. The Weak Gravity Conjecture is known to constrain the simplest case in which a single compact axion descends from a gauge field in an extra dimension. We argue that the Weak Gravity Conjecture also constrains a variety of theories of multiple compact axions including N-flation and some alignment models. We show that other alignment models entail surprising consequences for how the mass spectrum of the theory varies across the axion moduli space, and hence can be excluded if further conjectures hold. In every case that we consider, plausible assumptions lead to field ranges that cannot be parametrically larger than $M_{\rm Pl}$. Our results are strongly suggestive of a general inconsistency in models of large-field inflation based on compact axions, and possibly of a more general principle forbidding super-Planckian field ranges.
\end{abstract}

\section{Introduction}

If a significant primordial gravitational wave signal is detected in any near-future experiment, it will imply that the inflaton traversed a distance in field space larger than $M_{\rm Pl}$. This is the famous Lyth bound~\cite{Lyth:1996im} (for further refinements, see~\cite{Easther:2006qu,Baumann:2011ws,Antusch:2014cpa,Bramante:2014rva}). On first contemplating super-Planckian field ranges, an effective field theorist will tend to feel some discomfort, being inclined to write down the most general effective Lagrangian,
\be
{\cal L} = -  \frac{1}{2} m^2 \phi^2 + \frac{1}{2} \partial_\mu \phi \partial^\mu \phi  - \frac{\lambda}{4!} \phi^4 + \frac{c}{M_{\rm Pl}^2} \phi^2 \partial_\mu \phi \partial^\mu \phi   - \frac{\lambda_6}{M_{\rm Pl}^2} \phi^6 - \ldots.
\ee
For order-one values of the coefficients of high-dimension operators in $V(\phi)$, the potential will oscillate wildly over super-Planckian field ranges and spoil inflation. However, on further reflection, one realizes that a shift symmetry $\phi \to \phi + a$ would forbid every term in the potential, and thus a theory with a single dominant source of shift symmetry breaking can provide a technically natural approach to super-Planckian field ranges: we would predict that for some $f$, $\lambda \sim m^2/f^2$, $\lambda_6/M_{\rm Pl}^2 \sim m^2/f^4$, and so on. We could accommodate this shift symmetry in our effective field theory by parametrizing its breaking with a spurion and building a Lagrangian out of fields that nonlinearly realize the symmetry. The good behavior of the potential is enforced by the dominance of a single spurion. In fact, many physicists readily accept the axion as a potential solution to the strong CP problem, in spite of its severe Planck-suppressed operator problem~\cite{Kamionkowski:1992mf,Holman:1992us,Kallosh:1995hi}. To explain the very tight experimental bound on the effective theta angle in our universe, a theory of a generic pseudo-Goldstone axion must forbid high-dimension Planck-suppressed Peccei-Quinn-violating operators, so that QCD instanton effects provide the leading shift-symmetry-breaking spurion by several orders of magnitude. If the axion is a compact field (or equivalently, has an exact gauged discrete shift symmetry, $a \to a + 2\pi f$), then only exponentially small instanton effects can break the symmetry and it may be relatively easy to enforce single-spurion dominance. This may be realized by string theory axions, for example~\cite{Conlon:2006tq,Svrcek:2006yi}. The similarities in the need for a good shift symmetry to solve the strong CP problem and for large-field inflation have motivated concrete models of inflation which are natural from the effective field theory viewpoint~\cite{Freese:1990rb,Banks:1995dp,ArkaniHamed:2003wu,Kim:2004rp,Dimopoulos:2005ac,Silverstein:2008sg,McAllister:2008hb,Berg:2009tg,Germani:2010hd,Kaloper:2011jz,Kaloper:2014zba,Burgess:2014tja,Bachlechner:2014hsa,Csaki:2014bua,Furuuchi:2014cwa,Harigaya:2014rga,Higaki:2014mwa,Shiu:2015xda}. (Among these, the idea of $N$-flation builds on earlier work on assisted inflation, which was first studied in the context of exponential rather than axion-like potentials and then was generalized \cite{Liddle:1998jc, Copeland:1999cs, Mazumdar:2001mm, Jokinen:2004bp}. Another variation is $M$-flation \cite{Ashoorioon:2009wa,Ashoorioon:2011ki,Ashoorioon:2014jja}.)

While effective field theorists can find no dramatic problem with large-field inflation, they may retain some skepticism about the existence of UV completions. It has been suggested that quantum gravity will impose more severe constraints than EFT. Axion fields with super-Planckian decay constants appear to be rare in string theory vacua~\cite{Banks:2003sx,Rudelius:2014wla}, which may be a suggestive hint of a more general principle. It is generally believed that the set of quantum gravity theories is discrete (at least for nonsupersymmetric theories, where subtleties of continuous moduli spaces do not arise). This means that many apparently sensible effective field theories are actually in the swampland of theories that cannot be consistently coupled to gravity~\cite{Vafa:2005ui,ArkaniHamed:2006dz,Ooguri:2006in,Douglas:2005hq}. At this point we have relatively few guidelines for how to judge that a theory is in the swampland, but the Weak Gravity Conjecture (WGC) is among the sharpest and most powerful and well-motivated~\cite{ArkaniHamed:2006dz,Kats:2006xp,Banks:2006mm,Cheung:2014vva,Cheung:2014ega}. 

The WGC asserts that any theory containing both gravity and a massless abelian gauge field should have a charged particle in the spectrum whose mass is less than its charge in Planck units. To be precise, we require $m < \sqrt{2} q e M_{\rm Pl}$ in a four-dimensional theory in which gravitons and photons are the only massless particles. The motivation is to avoid having a plethora of exactly stable extremal black hole states, which are potentially problematic~\cite{Susskind:1995da}. This leads to:\\

\noindent
{\bf The Weak Gravity Conjecture (WGC)}: For any large, semiclassical, nearly-extremal black hole, there exists a state in the theory whose mass is small enough relative to its charge that the black hole can move away from extremality by emitting this state.\\

\noindent
For a single $U(1)$ gauge group, this implies that there is a state satisfying $q/m \ge z_0$ for an appropriate constant $z_0$.
In the case of multiple $U(1)$s, the condition is that the convex hull of the charge-to-mass vectors $\vec z = \vec q / m$ of the kinematically available charged states must contain the ball of radius $z_0$~\cite{Cheung:2014vva}. The is purely a kinematic requirement: black holes should be able to decay. 

From the very beginning, the WGC was claimed to rule out the theory of extranatural inflation~\cite{ArkaniHamed:2003wu}. Recently, there has been renewed interest in how the WGC can constrain large-field inflation~\cite{Rudelius:2014wla,Bachlechner:2014gfa,delaFuente:2014aca,Rudelius:2015xta,Montero:2015ofa,Brown:2015iha,Bachlechner:2015qja,Hebecker:2015rya,Brown:2015lia,Junghans:2015hba}. The essential idea is that theories of axions with good shift symmetries often obtain four-dimensional axion fields by dimensional reduction of higher-rank $p$-form fields, which are constrained by WGC arguments. 

However, such arguments are not without subtlety. The WGC is a rather weak statement at low energies, since large black holes could decay to states out of the reach of low-energy effective field theory. Given an effective field theory with cutoff $\Lambda$, the lightest semiclassical black holes have mass of order $M_{\rm Pl}^2/\Lambda$, so one could imagine that the conjecture is satisfied by states with mass between $\Lambda$ and $M_{\rm Pl}^2/\Lambda$ that cannot be studied without a full theory of quantum gravity. Moreover, extremal black holes can also satisfy the WGC, provided that the subleading corrections to the extremality bound have the correct sign~\cite{ArkaniHamed:2006dz,Kats:2006xp}. Nonetheless, the magnetic form of the WGC does have important consequences for the low-energy effective field theory and for inflation, which we review in~\S\ref{subsec:WGCreview}.

As effective field theorists, we might want to impose the stronger constraint that black hole decay can be described in the low energy effective field theory. That is, we might want to limit our attention to theories in which we can positively assert that black holes decay, reasoning that any theory which violates this assumption lies outside theoretical control in the absence of a full, quantum gravity description. This suggests a variant of the WGC:\\

\noindent
{\bf The Effective Weak Gravity Conjecture (EWGC)}: The state which satisfies the weak gravity conjecture should be describable in the low-energy effective field theory.\\

\noindent
Indeed, we usually imagine black holes decaying to particles, hence the EWGC is sometimes implicit in discussions of the WGC. However, we emphasize that it amounts to a further assumption, which however is much weaker than the ``strong form'' of the WGC proposed in~\cite{ArkaniHamed:2006dz}, which we discuss in~\S\ref{sec:strongforms}, but which plays no role in our arguments. By contrast, the EWGC plays an important role in some---but not all---of our arguments.

(We note in passing that the EWGC is implied by the much-stronger ``lattice WGC''~\cite{Heidenreich:2015nta}---discussed briefly in section~\ref{sec:LWGC}---whenever there is \emph{any} charged particle that can be described in the low-energy effective field theory.)

Unlike the WGC, the EWGC is not directly motivated by the problem of remnants, since Planck-scale states satisfying the WGC can address this issue. If correct, the EWGC may stem from a dynamical version of the (kinematic) WGC. For instance, one variation on the Third Law of Thermodynamics in the black hole context could be that nearly-extremal black holes should not spontaneously move closer to extremality.
To realize the implications of this statement, 
we note that black holes with sufficient charge do not decay predominantly by Hawking radiation. Initially, since Hawking radiation carries very little charge away, they decay \emph{towards} extremality \cite{Hiscock:1990ex}. The temperature of a Reissner-Nordstr\"om black hole goes to zero as the black hole approaches extremality, shutting off the Hawking radiation, but charged particles satisfying the WGC can still be emitted through an effect that is similar to Schwinger pair-production near the horizon~\cite{Gibbons:1975kk,Schumacher:1985zz,Khriplovich:1999gm}, which eventually becomes the dominant decay channel.

The effect may also be understood as a Breitenlohner--Freedman-type instability of charged particles in the near-horizon geometry~\cite{Denef:2009tp,Chen:2012zn,Chen:2014yfa}. All that is required is that at least one such particle exist; despite statements to the contrary in \cite{Banks:2006mm}, it need not be the lightest charged particle, as lighter particles violating the WGC inequality are not emitted. Conversely, if there are no charged particles in the low-energy effective field theory satisfying WGC, we expect that the black hole continues to approach extremality, as pair production cannot occur whereas the emission of long string states or fission of the black hole into smaller black holes should be a very slow process in comparison to Hawking radiation.\footnote{In principle, when the black hole is very close to extremality, the temperature of the Hawking radiation may become low enough that these strongly suppressed processes are competitive. However, we expect that for a large black hole with a weakly curved horizon, this transition occurs \emph{exponentially} close to extremality, hence it may not even be visible in the thermodynamic limit. Moreover, it's not clear that these processes are effective at discharging the black hole, since their rates are typically not under theoretical control.} Thus, the EWGC may be motivated by thermodynamic considerations.

As this paper was being finalized, we learned of work on the possibility that the WGC is satisfied by states that are not captured in the low-energy effective field theory \cite{Cornell2015}. Even if the EWGC does not hold in every consistent theory, the theories in which it is violated have the unusual property that the decay of large, semiclassical black holes cannot be described semi-classically. This suggests that the naive ``low energy effective field theory'' is not a completely reliable description of the full theory at low energies, and in particular any conclusions that we draw about inflation based solely on this class of theories may not be reliable. 

For the same reasons of theoretical control, in this paper we focus on the case where the low-energy effective field theory is a weakly-coupled abelian gauge theory, containing electrically charged particles light enough to discharge subextremal electrically charged black holes and semiclassical (solitonic) monopoles light enough to discharge subextremal magnetically charged black holes. This assumption could be circumvented in examples where a more sophisticated field theory description of the charged particles is available, e.g.\ in cases where the abelian gauge theory arises from Higgsing a non-Abelian gauge group and the monopoles originate from ``hedgehog'' configurations in the parent theory. However, we see no reason to expect that such theories will evade our constraints, hence for simplicity we defer consideration of them to future work.

\medskip

Our goal in this paper is to give a critical assessment of the state of large-field inflation in light of the WGC. We focus on scenarios with compact axion fields, leaving noncompact models of axion monodromy for future consideration (though some of our remarks may extend to such models). We find that arguments against approximately isotropic models of $N$-flation \cite{Dimopoulos:2005ac} and kinetic alignment \cite{Kim:2004rp,Bachlechner:2014hsa} are robust. The most difficult scenario to rule out arises from loopholes pointed out by de la Fuente, Saraswat, and Sundrum \cite{delaFuente:2014aca,Prashant}. We present conjectured bounds on this scenario that depend on the way that the mass spectrum shifts as the axion VEVs are varied. In every case that we find a bound, the parametrics {\em precisely} compensates for any possible enhancement and determines that the field range is bounded above by $M_{\rm Pl}$ (times at most an order-one number).

The organization of this paper is as follows. In section \ref{sec:preliminaries}, we review extranatural inflation~\cite{ArkaniHamed:2003wu}, the Weak Gravity Conjecture and its requirement of a low UV cutoff \cite{ArkaniHamed:2006dz}, and arguments against large-field inflation with a single axion from both the electric \cite{ArkaniHamed:2006dz} and magnetic \cite{delaFuente:2014aca} points of view. In section \ref{sec:multiaxion}, we present arguments for how the magnetic form of the Weak Gravity Conjecture excludes the simplest models of $N$-flation \cite{Dimopoulos:2005ac} and models based on the Kim--Nilles--Peloso alignment mechanism \cite{Kim:2004rp} when it is realized through the structure of kinetic mixing \cite{Bachlechner:2014hsa}. These arguments do not address the case of decay constant alignment where the leading instanton effects arise from highly aligned electric charges in a basis in which the magnetic charges are not aligned \cite{delaFuente:2014aca}. In section \ref{sec:kkmonodromy}, we discuss how the spectrum of modes for extranatural inflation varies while traversing the axion moduli space. We argue that the models that evade our earlier arguments involve spectra with surprising features that one must accept in order to realize large-field inflation. We claim that, in the presence of particles of large charge, the 5d effective field theory breaks down in some regions of the moduli space if the compactification radius is not significantly smaller than the UV cutoff. This motivates the Single-EFT Consistency Criterion (SECC), which demands that any Kaluza-Klein mode which is light in some region of moduli space should have mass below the UV cutoff throughout the moduli space. Equivalently, it requires that the axion shift symmetry $\theta \to \theta + 2\pi$ can be understood as a large gauge transformation within the domain of validity of the UV completion. We emphasize that the SECC can be motivated within the 5d effective field theory in the presence of a Wilson line, not just the dimensionally reduced viewpoint. (In appendix \ref{sec:lattice} we further illustrate the SECC using a gauge-invariant lattice regulator.) A second new conjecture, the Extended Weak Gravity Conjecture, requires that the spectrum of particles should satisfy the WGC bound at every stationary point in moduli space (including local maxima). In section \ref{sec:decayconstantalignment}, we show that the loopholes raised by \cite{delaFuente:2014aca} violate our stronger conjectures. This completes our discussion of bounds on inflation from the WGC.

Sections \ref{sec:entropybounds} and \ref{sec:strongforms} address prior claims to have constrained large-field inflation models using either entropy bounds \cite{Kaloper:1999tt,Conlon:2012tz,Boubekeur:2013kga} or hypothetical strong forms of the WGC \cite{Brown:2015iha}. These sections exist to place our paper in context and can be freely skipped. We claim that the entropy bound arguments in the literature rest on overly strong assumptions, and that the ``strong form'' of the WGC is ambiguous when applied to multiple $U(1)$s. We emphasize that we have never used such a strong form in this paper, and that we view the issues of Kaluza-Klein mode monodromy to be the most important new requirements we have used.

We conclude in section \ref{sec:conclude} with a discussion of our view of what remains to be done to place our conjectured requirements on a sounder footing. We believe that further progress along these lines can either thoroughly exclude parametrically large field inflation or identify special theories that satisfy all consistency requirements of quantum gravity.

\section{Preliminaries}
\label{sec:preliminaries}

In this section we review the concept of extranatural inflation, the Weak Gravity Conjecture and its implications for the UV cutoff of a theory, and how the WGC rules out models of inflation driven by a single axion field. Along the way we make a few small remarks not present in the existing literature, but readers thoroughly familiar with the WGC can skip to the next section for our new results.

\subsection{Axions from extra dimensions}

In this paper we will focus on extranatural inflation models~\cite{ArkaniHamed:2003wu} in which the axion field arises by reducing a $p$-form gauge field ($p \geq 1$) on a $p$-dimensional cycle within a compactification manifold. String theory axions \cite{Conlon:2006tq,Svrcek:2006yi} share this feature with simple phenomenological models. It seems plausible that any consistent theory of a compact axion field coupled to quantum gravity can be viewed, in some duality frame, as a member of this class, so we view the restriction to extranatural models as a mild assumption.

For example, consider the case of an ordinary (1-form) gauge symmetry. If the compactification manifold contains a circle, we can define an axion-like field in four dimensions from the integral around this circle:
\be
\theta(x) = \oint_0^R dx^5 A_5(x,x^5).
\ee
Large gauge transformations of $A_5$ lead to $\theta \to \theta + 2\pi n$ (where $n \in \mathbb{Z}$) representing a (gauged) discrete shift symmetry and signifying that the axion field is compact. The same statement holds for axions obtained from higher-rank $p$-forms, where there is always the freedom to perform a large gauge transformation shifting the field by an integral multiple of the volume form of the $p$-cycle we integrate over.

Beginning from a 5d action $\int d^5 x (\frac{1}{2} M_5^3 {\cal R} - \frac{1}{4 e_5^2} F_{\mu\nu}^2)$, we obtain a 4d Planck scale set by $M_{\rm Pl}^2 \equiv M_4^2 = 2\pi R M_5^3$, gauge coupling $1/e^2 = 2\pi R/e_5^2$, and an axion decay constant set by
\be
{\cal L}_{\rm axion} = \int d^4 x \frac{1}{2} f^2 \partial_\mu \theta \partial^\mu \theta,~~{\rm where}~~f^2 = \frac{1}{2\pi R\, e_5^2} = \left(\frac{1}{2\pi R\, e}\right)^2.
\ee
In much of the paper, we will focus on this case, where a four-dimensional axion arises from the dimensional reduction of a one-form gauge field in five dimensions.
Nonetheless, most of our arguments generalize straightforwardly to axions originating from $p$-form gauge fields in $D$ dimensions.

\subsection{Weak Gravity Conjecture} \label{subsec:WGCreview}

A useful formulation of the Weak Gravity Conjecture is that any near-extremal charged black hole should be able to move away from extremality by emitting a charged particle. For a $D$-dimensional theory containing a $p$-form field with coupling $g_p$ also coupled to a massless canonically normalized dilaton $\varphi$ through its kinetic term $\propto e^{-\alpha \varphi \sqrt{16 \pi G}} F_p^2$, the WGC asserts that there should exist a $(p-1)$-brane with charge $q$ under the $p$-form and tension $T$ satisfying the inequality 
\be
8 \pi G \left[\frac{\alpha^2}{2} + \frac{p(D-p-2)}{D-2}\right] T^2 \leq g_{p}^2 q^2.
\ee
We comment on the detailed $\alpha$, $p$, and $D$ dependence of this expression in a separate paper \cite{Heidenreich:2015nta}. 

For a $U(1)$ gauge theory in four dimensions with coupling constant $e$, this has the immediate implication that there should exist an electrically charged particle of mass $m$ and charge $q$ satisfying $m < \sqrt{2} e q M_{\rm Pl}$ and a magnetic monopole of mass $m_{\rm mon}$ and charge $q_m$ satisfying $m_{\rm mon} < \sqrt{2} q_m M_{\rm Pl}/e$. As pointed out in \cite{ArkaniHamed:2006dz}, the existence of a monopole satisfying such a bound implies that weakly-coupled gauge theories coupled to gravity must have a UV cutoff that is parametrically below the Planck scale. The self-energy of the monopole from integrating its classical magnetic field down to a distance $r_{\rm min} \sim \Lambda^{-1}$ is linearly divergent: $\delta m_{\rm mon} \sim \frac{q_m^2}{e^2} \Lambda$. The precise coefficient is not essential because we make a naturalness argument, that $m_{\rm mon} \simgt \delta m_{\rm mon}$ (in the absence of fine-tuning), which like all naturalness arguments leaves the precise order-one coefficient in the bound as a matter of taste. Combining the naturalness bound with the magnetic WGC, we learn that
\be
\Lambda \simlt \frac{e}{q_m} M_{\rm Pl}.
\ee
In other words, the UV cutoff of the theory should be below $e M_{\rm Pl}$, and if the magnetic WGC is satisfied by a monopole of large charge the UV cutoff will be even lower. We emphasize that this magnetic form of the WGC, from~\cite{ArkaniHamed:2006dz}, does not depend on the EWGC, and our arguments based on it are the most robust.

Notice that although one could attempt to apply the self-energy argument also to the electric field of electrically charged particles, such an integral must be cut off at the Compton wavelength $m^{-1}$ of the field. As is familiar from QED, the self-energy of the electron is a small correction to its mass. The argument that monopoles can bound the cutoff $\Lambda$ relies on the fact that the magnetic coupling is strong, the monopole is a solitonic object with mass above the cutoff, and that electric and magnetic charges are mutually nonlocal, so a theory containing both necessarily has a fundamental cutoff. The interpretation of the quantity $\Lambda$ that we are bounding is the scale at which it is no longer appropriate to treat the theory as a weakly-coupled, local effective field theory of a $U(1)$ gauge boson. This may be because $U(1)$ embeds into a nonabelian group which has $W$ bosons at the scale $\Lambda$, or may be due to more fundamental new physics such as the string scale or the higher-dimensional Planck scale.

\subsection{The list of ingredients}

In this paper there are four ingredients that will play a role in our arguments:
\begin{enumerate}
\item {\bf The instanton effects.} We are dealing with a set of axions that have a potential of the form $\sum_i c_i \cos(\sum_{j} Q_{ij} \theta_j)$. The charge matrix $Q$ is typically determined by a set of wrapped worldvolumes of electrically charged particles in extranatural inflation theories, but could originate from other physics. 
\item {\bf The electric charges satisfying the WGC.} A set of electrically charged particles exist that satisfy the constraints demanded by Weak Gravity. That these may not be the {\em same} charged particles contributing the dominant instanton effects is one source of loopholes to the simplest arguments.
\item {\bf The magnetic charges satisfying the WGC.} A set of magnetically charged particles exist that satisfy the constraints of Weak Gravity. These particles give us crucial access to information about the UV cutoff of the theory and are the central elements of many of our arguments.
\item {\bf The kinetic mixing matrix.} Our axions (or the higher-dimensional gauge fields they originated as) in general mix with each other through a matrix $K$.
\end{enumerate}

A number of models in the literature rely on ``alignment,'' following the work of Kim, Nilles, and Peloso \cite{Kim:2004rp}. A general alignment model is one in which all four of our ingredients can freely vary. In this paper we address two {\em physically distinct} special cases of alignment, which are not simply the same idea in a different basis. We will use the term {\em kinetic alignment} to refer to a scenario in which the electric charges contributing the dominant instanton effects (our first ingredient) and the magnetic charges satisfying the WGC (our third ingredient) are {\em both} simple (small integers in the charge lattice) in a basis where $K$ is arbitrary. This addresses models like \cite{Bachlechner:2014hsa}, though we make the additional assumption of simple magnetic charges. On the other hand, we could consider a model in which the kinetic term is simple but the charge matrix of the instanton effects and the magnetic charges satisfying the WGC are not (so that, for instance, there are large numbers appearing in the instanton matrix while the magnetic charges are simple). This is the scenario of \cite{delaFuente:2014aca}, which we refer to as {\em decay constant alignment}. Given this scenario, we could always redefine our basis to make the electric charges simple while scrambling the kinetic matrix. However, we would then have unusual charge assignments for the magnetic monopoles, and we require a different physical argument than the one we applied to the scenario we referred to as kinetic alignment. Although we only obtain bounds on these two special cases, we expect that a combination of the ideas used in these bounds can exclude the general case.

To fix conventions for the electric charge in the extranatural case, we write the coupling of a charged particle to the gauge potential as
\be
S = Q_a \int_P A^a\,,
\ee
where the integral is over the worldline of the particle, and $a$ indexes the different gauge fields in the case of multiple $U(1)$'s. This leads to an axion potential of the form
\be
V = V_0 \sum_n c_n \cos (n Q_a \theta^a)\,,
\ee
in the dimensionally reduced theory, where $\theta^a = \oint A^a$.
Likewise, we define the magnetic charge enclosed in a spatial region $\Sigma$ as
\be
\tilde{Q}^a \equiv \frac{1}{2 \pi} \int_{\partial \Sigma} F^a\,,
\ee
where $F^a = d A^a$. The Dirac quantization condition is
\be
\tilde{Q}^a Q_a \in \mathbb{Z} \,.
\ee
Unless otherwise specified, we always work in a basis where $\tilde{Q}^a$ and $Q_a$ are integrally quantized.

\subsection{WGC and single-axion inflation: electric argument} \label{subsec:elecWGCinflation}

An argument that the WGC excludes extranatural inflation with super-Planckian decay constants was given already in the original paper \cite{ArkaniHamed:2006dz}. The electric WGC in 5 dimensions tells us that a charged particle exists with mass $m < \sqrt{3/2} e_5 q M_5^{3/2}$. Upon reducing to four dimensions, this implies that the charged particle contributes an instanton action
\be
S_{\rm inst} = 2\pi R m <  \sqrt{\frac{3}{2}}  \frac{q M_{\rm Pl}}{f}.
\label{Sbound}
\ee
Let us assume that the same instanton generates the axion potential.
If we want to focus on large instanton actions so that higher-order corrections to the potential are exponentially suppressed, this shows that we require $f/q \simlt M_{\rm Pl}$, where $f/q$ is the field range determined by this potential.

As recently emphasized in \cite{delaFuente:2014aca,Prashant}, this argument is not completely convincing because there is a ``small action loophole'': the prefactor in front of higher-instanton terms can suppress them even if the instanton action $S \ll 1$. Direct calculation confirms that the potential is a sum of terms proportional to $\exp(-2\pi n m R)\cos(n a/f)$ with coefficients decreasing with powers of $n$~\cite{Hosotani:1983xw,Cheng:2002iz,ArkaniHamed:2003wu}. Even at $m = 0$ where the instanton action gives no suppression, the $n^{\rm th}$ term in the sum has a $1/n^5$ suppression which is enough to safely allow inflation. A useful way to understand this power law suppression is to write the contribution of winding number $n$ in terms of the 5d Green's function for charged particle propagation $n$ times around the circle (see Appendix A of \cite{ArkaniHamed:2007gg}), in which case the power law is just the usual cost of propagating a massless field over the long distance $2\pi R n$.

One can partially constrain this small action loophole by demanding that the convex hull condition should be satisfied for a 4d theory that includes both the usual $U(1)$ and a Kaluza-Klein $U(1)$.  As shown in \cite{Heidenreich:2015nta}, this convex hull condition yields the inequality:
\begin{equation}
m_0 R \geq \frac{1}{2 z_0 \left(z_0^2-1\right)^{1/2}},
\label{bound2}
\end{equation}
where
\begin{equation}
z_0 = \sqrt{\frac{3}{2}} \frac{e  q M_{\rm Pl}}{m_0} = \sqrt{\frac{3}{2}} \frac{ q M_{\rm Pl} }{2 \pi f m_0 R}.
\end{equation}
Here $m_0$ is the mass of the particle in 5d, and we have turned off the dilaton in the 5d theory.  The convex hull condition for the 5d theory just enforces $z_0 \geq 1$, which is equivalent to (\ref{Sbound}).
For $q=1$, we see that $f \sim (m_0 R)^{-1}$, so the maximal value of the decay constant grows inversely with the instanton action $S =  2 \pi m_0 R$ in the limit $S \rightarrow 0$.  However, imposing the stronger condition (\ref{bound2}) and looking at the $m_0 R \rightarrow 0$ limit, we find
\begin{equation}
f^2 \leq \frac{3 M_{\rm Pl}^2 }{(2 \pi)^2 m_0 R}.
\end{equation}
This tells us that the maximal allowed value of $f$ grows like $S^{-1/2}$, rather than the na\"ively expected $S^{-1}$.  Thus, the weak gravity bound on axion decay constants in the context of extranatural inflation is stronger than expected, but it is not strong enough to close the small action loophole---one may achieve a super-Planckian decay constant by taking $S$ small without violating the electric form of the WGC.

\subsection{WGC and single-axion inflation: magnetic argument}
\label{subsec:magneticargument}

The small action loophole led \cite{delaFuente:2014aca} to propose a different argument for how the WGC can exclude single-axion extranatural inflation. Starting with the UV cutoff $\Lambda \simlt e M_{\rm Pl}$, they demand that the size of the compactification manifold be larger than $\Lambda^{-1}$. Then
\be
1 \simlt 2\pi R \Lambda \simlt 2\pi R\, e M_{\rm Pl} = \frac{M_{\rm Pl}}{f}.
\ee
In this way we obtain the constraint $f \simlt M_{\rm Pl}$, completely independent of the size of the leading instanton effect or which charged particle generates it. Notice that because the cutoff is $\Lambda < eM_{\rm Pl}/q_m$ when the monopole satisfying magnetic WGC has charge $q_m$ and a particle of electric charge $q$ leads to an instanton proportional to $\cos(q a/f)$, choosing nonminimal charge assignments anywhere in the argument only makes the bound stronger.

\section{Weak Gravity Conjecture and multi-axion models}
\label{sec:multiaxion}

In this section we will extend the magnetic argument against single-axion inflation from section \ref{subsec:magneticargument} to more general scenarios with multiple axion fields. We will first tackle the simplest case of $N$-flation with the strong assumption of diagonal kinetic terms and minimal charge assignments. Then we relax our assumptions to consider alignment models.

\subsection{Warmup: diagonal N-flation} \label{sec:diagonalwarmup}

Suppose that we have $N$ $U(1)$ gauge fields $A_i$ with couplings $e_i$ and that the kinetic mixing among them is negligible. We will also assume that in this basis the charge lattice simply consists of integer electric or magnetic charges under each $U(1)$. These are strong simplifying assumptions, but provide a useful starting point. The Weak Gravity Conjecture applied to each gauge field separately implies the existence of electrically and magnetically charged particles satisfying certain bounds. But the constraint for the set of $N$ fields is actually stronger than for any individual field: if we marginally saturate the bound for each $U(1)$ by postulating a magnetic monopole with charge $q_m = 1$ and mass $\sqrt{2} M_{\rm Pl}/e_i$, for example, then a nearly-extremal black hole with large and equal charges under every $U(1)$ will not be able to decay. This is because the extremality bound for a black hole charged under multiple groups depends not on the sum of the charges but on the charges added in quadrature: for a magnetically charged black hole in four dimensions, the bound is $Q_{\rm eff} \equiv \sqrt{Q_1^2/e_1^2 + \ldots + Q_N^2/e_N^2} < M_{\rm BH}/(\sqrt{2} M_{\rm Pl})$.

Consider extranatural $N$-flation that moves along the diagonal in each axion direction, attempting to obtain an effective decay constant
\be
f_{\rm eff}^2 = f_1^2 + f_2^2 + \ldots + f_N^2 = \left(\frac{1}{2\pi R}\right)^2 \left(\frac{1}{e_1^2} + \ldots + \frac{1}{e_N^2}\right).
\label{eq:feffsqdiagonal}
\ee
Notice that this is the appropriate expression under the assumption of no kinetic mixing {\em and} the further assumption that the dominant instanton effects give rise to a potential of the form $\sum c_i \cos(a_i/f_i)$, as would be generated for example from wrapped worldlines of electrically charged particles of charge 1 under each gauge group. More general instantons can lead to alignment phenomena in which inflation winds around one direction in axion space multiple times. We will return to such a possibility later. 

The linear combination of $1/e_i^2$ factors appearing on the right-hand side of (\ref{eq:feffsqdiagonal}) is precisely what appears in the extremality bound for a magnetically charged black hole with equal charge $Q$ under all $N$ gauge groups. Let us build some intuition by considering ways that such a diagonally magnetically charged black hole could decay:

\begin{itemize}

\item It could emit a monopole of diagonal charge $(q,q,\ldots q)$, so that its charge-to-mass vector points in the same direction after the emission but is now shorter (we take $Q > q > 0$). In this case, the problem essentially reduces to the one-field case. The self-energy of the monopole imposes $m_{\rm mon} \simgt q^2 \sum_i \frac{1}{e_i^2} \Lambda$, while the condition that the black hole moves away from extremality imposes that $m_{\rm mon}^2 \simlt q^2 \sum_i \frac{1}{e_i^2} M_{\rm Pl}^2$. These conditions together with $2\pi R \Lambda \simgt 1$ require that $f_{\rm eff} \simlt M_{\rm Pl}/q$.

\item It could emit a monopole charged under a single gauge group. Suppose it emits a particle with mass $m_1$ and charges $(q_1, 0, \ldots, 0)$. The self-energy constraint leads to $m_1 \simgt q_1^2 \Lambda/e_1^2$. If the diagonally-charged black hole emits this particle, its effective charge decreases only by (expanding the square root) $-\Delta Q_{\rm eff} \approx Q q_1/(e_1^2 Q_{\rm eff}) = (q_1/e_1) (f_1/f_{\rm eff})$. As a result, the condition that the monopole can be emitted is no longer $m_1 < \sqrt{2}  q_1/e_1 M_{\rm Pl}$ but the stronger condition $m_1 < \sqrt{2} q_1/e_1 (f_1/f_{\rm eff}) M_{\rm Pl}$. This leads to $f_{\rm eff} \simlt M_{\rm Pl}/q_1$.

\item Now consider the general case in which the monopole emitted has mass $m$ and charges $(q_1, \ldots q_N)$. For the black hole to move away from extremality we first require that $Q_{\rm eff}$ decreases, so that $\sum q_i/e_i^2 > 0$. A straightforward generalization of the previous argument leads to a bound
\be
f_{\rm eff} \simlt M_{\rm Pl} \frac{\sum_i q_i/e_i^2}{\sum_i q_i^2/e_i^2} \simlt M_{\rm Pl}.
\ee
The last step follows because charge quantization demands that $q_i^2 > |q_i|$.

\end{itemize}

These arguments give a suggestive hint of how scenarios with multiple axions can be more strongly constrained by the Weak Gravity Conjecture in a manner that precisely compensates for the expected gain in field range. However, we have made a strong simplifying assumption that the electric charges leading to dominant instanton effects are simple in the same basis that the gauge field kinetic term is diagonal. We will now explore the constraints imposed by the WGC if we relax this assumption.

\subsection{Magnetic WGC and kinetic alignment}\label{sec:magwgckineticalignment}

Consider the case of a general kinetic matrix for the gauge fields:
\be
-\frac{1}{4} K_{ij} F^i_{\mu \nu} F^{j\mu\nu}.
\ee
Assume that we are working in a basis in which there are $N$ magnetic monopoles that satisfy the magnetic WGC and have unit charges $(1,0,\ldots,0), (0,1,\ldots,0), \ldots (0,0,\ldots 1)$. We can choose a different basis to diagonalize the kinetic terms:
\be
K = O D O^T,
\label{KD}
\ee
where $O$ is an orthogonal matrix and $D = {\rm diag}(1/g_1^2, \ldots 1/g_N^2)$ is a diagonal matrix. Without loss of generality we can choose
\be
g_{\rm min}^2 \equiv g_1^2 \leq g_2^2 \leq \ldots \leq g_N^2 \equiv g_{\rm max}^2.
\ee
In this basis, the $i^{\rm th}$ monopole has charge assignments ${\vec o}_i$ that can be read off from the matrix $O$. The self-energy of this monopole gives us an inequality relating the UV cutoff $\Lambda$ and the monopole mass $m_i$:
\be
\left(\sum_j \frac{o_{ij}^2}{g_j^2} \right) \Lambda \simlt m_i.
\ee
In particular, if we sum over all $i$ and exploit orthogonality, we learn that
\be
\sum_j \frac{1}{g_j^2} \Lambda \simlt m_1 + \cdots + m_N \equiv m_{\rm tot}.
\ee

The magnetic WGC tells us that the convex hull of the charge-to-mass vectors $\pm {\vec z}_i$ of the $N$ monopoles contains the unit ball. We have
\be
({\vec z}_i)_j = \frac{o_{ij} M_{\rm Pl}}{g_j m_i}.
\ee
Since the ${\vec z}_i$ form a basis, the convex hull condition can be restated as the requirement that
for any coefficients $\alpha_1, \ldots, \alpha_N$,
\be
\left| \sum_i \alpha_i {\vec z}_i\right| \geq \sum_i \left|\alpha_i\right|.
\ee
Consider the choice $\alpha_i = \sigma_i m_i$, where $\sigma_i = \pm 1$ is a choice of sign. In this case the convex hull condition tells us that
\be
m_1 + \cdots + m_N \leq \sqrt{\sum_j \left(\sum_i \frac{\sigma_i o_{ij}}{g_j}\right)^2} M_{\rm Pl}.
\ee
Combining the convex hull condition with the constraint on the cutoff, we learn that for any set of sign choices $\sigma_i$
\be
\sum_j \frac{1}{g_j^2} \Lambda\simlt \sqrt{\sum_j \left[\frac{1}{g_j^2} \left(\sum_i \sigma_i o_{ij}\right)^2\right]} M_{\rm Pl}.
\label{sumeq}
\ee
There are $2^N$ choices of sign $\sigma_i$, some of which potentially provide much stronger bounds than others.

Consider the case where the largest eigenvalue completely dominates, so that we can drop all terms in the sum not proportional to $\frac{1}{g_1^2}$. In \cite{Bachlechner:2014hsa}, it was pointed out that the eigenvector with largest eigenvalue $1/g_1^2$ of a randomly chosen kinetic matrix $K_{ij}$ will almost certainly point close to a diagonal direction of the fundamental cube, e.g. $\sim (1,1,\ldots 1)/\sqrt{N}$. Since the diagonal of an $N$-dimensional cube has length $\sqrt{N}$, this implies (in the extranatural context with minimal instanton charges) an effective decay constant of $f_{\rm eff} \approx \sqrt{N}/(2 \pi R g_1)$ in the direction of largest eigenvalue. However, in this case, we can choose the signs $\sigma_i$ to alternate and nearly cancel so that the sum $\sum_i \sigma_i o_{i1} \sim \frac{1}{\sqrt{N}}$. This leads to an estimated bound
\be
\frac{1}{g_1^2} \Lambda \simlt \sqrt{\frac{1}{g_1^2 N}} M_{\rm Pl}~~\Rightarrow~~\Lambda \simlt \frac{g_1 M_{\rm Pl}}{\sqrt{N}}.
\ee
This bound is larger than the naive one-field version of the magnetic WGC by a factor of $\sqrt{N}$, so imposing that $\Lambda R \simlt 1$ precisely produces $f_{\rm eff} \simlt M_{\rm Pl}$.

It is possible to make a more general version of the argument that excludes any case in which one eigenvalue $g_1$ dominates the sums (\ref{sumeq}), without making an assumption about the eigenvector. The cutoff $\Lambda$ in this case obeys
\be
\Lambda \lesssim g_1 \left| \sum_i \sigma_{i} o_{i1} \right| M_{\rm Pl}.
\label{smallcutoffeq}
\ee
On the other hand, the 4d Lagrangian for the axions is given by
\be
\frac{1}{2(2 \pi R)^2} D_{ij} \partial_\mu\theta_i \partial^\mu\theta_j - \sum_i A_i e^{-S_i} \cos\left(\sum_j o_{ij} \theta_j \right),
\ee
where $D$ is the diagonal matrix of (\ref{KD}).  In the limit in which $1/g_1^2$ is much larger than the other eigenvalues of $D$, we may approximate the axion moduli space radius by considering only the field displacement in the direction of largest eigenvalue $|\Delta\theta_1|$.  The maximal displacement is given by the largest value of $|\theta_1|$ satisfying the conditions
$$
|o_{i1} \Delta\theta_1| \leq \pi,
$$
for all $i$.  Thus,
\be
|\Delta\theta_1|_{\rm max} = \frac{\pi}{{\rm Max}_i|o_{i1}|}.
\ee
Combining this with our previous bound (\ref{smallcutoffeq}) and setting $\Lambda R \simgt 1$, we get a bound on the radius of axion moduli space,
\be
r \simlt \frac{\Lambda}{2\pi g_1} |\Delta\theta_1|_{\rm max} \simlt \frac{1}{2} \left| \frac{ \sum_i \sigma_{i} o_{i1} }{{\rm Max}_i|o_{i1}|} \right| M_{\rm Pl}
\label{rboundsmallg1}
\ee

Finally, it is not hard to see that we can choose the signs $\sigma_i$ so that this fraction is smaller than $1$.  Order the $o_{i1}$'s in descending order of their magnitude.  Set $\sigma_1 = +1$.  Then, recursively define,
\be
   \sigma_k = \left\{
     \begin{array}{lr}
       {\rm sgn}(o_{k1}) & \mbox{ :     }~~~\sum_{i=1}^{k-1} \sigma_i o_{i1} < 0 \\
       -{\rm sgn}(o_{k1}) & \mbox{ :     }~~~\sum_{i=1}^{k-1} \sigma_i o_{i1} \geq 0
     \end{array}
   \right.
\ee
Since the $o_{k1}$s are decreasing in magnitude, the partial sum $|\sum_{i=1}^{k-1} \sigma_i o_{i1}|$ can never jump more than $|o_{11}|$ by adding a new term.  By picking the signs in this way, we ensure that we are always moving towards the origin, so the magnitude of the partial sums is necessarily decreasing with $k$.  Since the magnitude of the first partial sum is just $|o_{11}|$, we see that the full sum must be smaller in magnitude than $|o_{11}|$.  Thus, (\ref{rboundsmallg1}) gives
\be
r \simlt \frac{1}{2} M_{\rm Pl}.
\ee
This excludes any kinetic alignment model with a single dominant large eigenvalue, again under the assumption that the instanton effects are controlled by minimal electric charges in the same basis for which the magnetic monopoles satisfying the WGC have minimal charge. A {\em parametric} violation of this assumption, such as instanton effects that are highly aligned, can evade our arguments. We will discuss such a case in section \ref{sec:decayconstantalignment}. The assumption of single-eigenvalue dominance, on the other hand, is made only for simplicity. It is straightforward to check in the two-axion case that the bound holds for completely arbitrary eigenvalues.  Furthermore, simple numerical studies in which the kinetic matrix is chosen from a Wishart distribution, $K_{ij} \sim W_{N}(\sigma^2,N)$, reveal that indeed the radius of moduli space decreases with increasing $N$.

\section{New conjectures on EFT over the moduli space}
\label{sec:kkmonodromy}

\subsection{Exploring the moduli space: masses and Kaluza--Klein reduction}

In this section we develop a new tool for constraining large-field axion models arising from extra dimensions, which opens an opportunity to obtain powerful constraints on models of axion monodromy. This approach relies, in part, on the nontrivial manner in which shift symmetries are realized in the effective theory. The potential energy in extranatural inflation (including string axion models) is a sum of cosine terms from instantons of various winding numbers, respecting an exact discrete shift symmetry. However, other terms in the effective theory preserve the shift symmetry in a less transparent way. Consider the case of a 4d axion obtained by dimensional reduction of a 5d 1-form gauge field, and suppose that in five dimensions there is a fermion $\Psi$ with charge $q$ under the gauge field. (The case of a charged scalar field is similar.) Its action is
\be
\int d^5 x \sqrt{-g} \left(i {\bar \Psi} \Gamma^M D_M \Psi + m_5 {\bar \Psi} \Psi + \frac{c}{\Lambda} D_M {\bar \Psi} D^M \Psi + \ldots\right), \label{eq:5dchargedaction}
\ee
where $D_M = \partial_M - i q A_M$, $\Lambda$ is the UV cutoff of the theory, and the dots represent various higher-dimension operators. The five Dirac matrices $\Gamma^M$ correspond to the usual 4d Dirac matrices together with $-i \gamma^5$. We emphasize that $\Lambda$ is the scale at which the local, 5d abelian gauge theory breaks down. In particular, we have no guarantee of five-dimensional locality holding at distances shorter than $\Lambda^{-1}$.

We study this theory on a background of ${\mathbb R}^{3,1} \times S^1$ with the fifth dimension having a periodic identification $y \sim y + 2\pi R$ with a background gauge field $A_5 = \frac{\theta}{2\pi R}$. Although fixing $A_5$ to be constant is a gauge choice, there is a gauge-invariant Wilson loop determined by $\theta$ which is well-defined modulo $2\pi$. The compactified theory contains a term
\be
\int d^5 x \sqrt{-g} \frac{q\theta}{2\pi R} {\bar \Psi} \Gamma^5 \Psi, \label{eq:effectivemassterm}
\ee
that we may think of as an effective mass (albeit one that depends on the spontaneous breaking of 5d Lorentz symmetry) which potentially decouples $\Psi$ from the effective theory if $\theta$ is large enough. This 5d term gives rise to a (CP-odd) mass term $\propto i \theta {\bar \psi} \gamma^5 \psi$ in the 4d theory, which can have important dynamical consequences when $\theta$ is large. At first glance, such a mass term constitutes a hard breaking of the shift symmetry for $\theta$---even of the {\em gauged} $\theta \to \theta + 2\pi$ symmetry! The resolution of this puzzle is that there is a monodromy in the Kaluza--Klein spectrum. Using the Kaluza-Klein decomposition $\Psi(x,y) =  \sum_{n = -\infty}^{\infty} \exp(i n y/R) \psi_n(x)/\sqrt{2\pi R}$, this action leads to a 4d effective theory
\be
{\cal L}_{\rm eff} =  \sum_{n = -\infty}^{\infty} \left(i {\bar \psi}_n \gamma^\mu D_\mu \psi_n + m_5 {\bar \psi}_n \psi_n + i \frac{n - \frac{q \theta}{2\pi}}{R} {\bar \psi_n} \gamma^5 \psi_n + \frac{c}{\Lambda} \left|\frac{n - \frac{q \theta}{2\pi}}{R}\right|^2 {\bar \psi}_n \psi_n+ \ldots \right). \label{eq:kkdecomposedL}
\ee
If we were to truncate this theory to a few low-lying modes, we would find a violation of the shift symmetry $\theta \to \theta + 2\pi$. But this symmetry is a large gauge transformation in the higher-dimensional UV completion, so it cannot be violated. Writing the EFT for all Kaluza--Klein modes makes the answer manifest. There is a monodromy effect that rearranges the spectrum; when $\theta \to \theta + 2\pi$, the mode with label $n$ acquires the same mass spectrum that the mode with label $n - q$ previously had. Because the derivative $\partial_5$ and the contribution of $A_5$ are always packaged together in a covariant derivative, this will be true of arbitrary higher-dimension operators as well.

Recall that for a Dirac fermion with mass term $m {\bar \psi} \psi + i \mu {\bar \psi} \gamma^5 \psi$, the physical mass is $\sqrt{m^2 + \mu^2}$. In particular, all 4d fields have mass larger than the 5d mass $m_5$.

\subsection{Consistency of a single EFT across axion moduli space}\label{sec:secc}

We have seen that the 5d theory compactified on a circle with a Wilson loop $\theta = \oint A_5 dx^5$ turned on has a spectrum that depends nontrivially on the value of $\theta$. Let us ask what happens when we move a large distance in moduli space. Tracking a single KK mode adiabatically as $\theta$ varies, we find that its CP-odd mass is shifted by
\be
\Delta m = \frac{q \Delta\theta}{2 \pi R}
\ee
In particular, if $\Delta \theta \simgt 2 \pi R \Lambda/q$, then a KK mode which is initially light acquires a large mass of order the cutoff $\Lambda$, and exits the effective theory. In fact, when we move this far in moduli space, the entire KK spectrum is shifted, so that the modes which were initially light are heavy, and modes initially above the cutoff are light. Since our description of the five-dimensional theory breaks down at $\Lambda$ (and in particular 5d locality may not hold above this scale), it is possible that in the process new physics can emerge from the cutoff and become light, ruining our effective description. Thus, if we wish to retain control of the KK spectrum, we should impose:
\beq
q\, \Delta \theta \simlt 2 \pi R \Lambda.
\label{eq:monodromyfieldrange}
\eeq
We emphasize that this is {\em not} a statement about the 4d effective theory cut off at the compactification scale, which obviously does not include Kaluza--Klein modes that may be important elsewhere in the moduli space. It is a statement about the 4d theory including a tower of weakly coupled modes all the way up to the cutoff $\Lambda$, which is fully equivalent to the 5d theory on the Wilson loop background. One point that we should emphasize is that the breakdown of effective field theory that are we discussing does {\em not} correspond to a violation of perturbative unitarity in high-energy scattering in 5d. The Wilson loop is gauge-invariant only when integrated over the full circle, so short-distance 5d scattering experiments do not detect it. Local scattering experiments are not the only way to detect a failure of EFT, however, and the KK mode spectrum is a physical observable that does so.

One clear instance in which a subtlety of this kind \emph{does not} arise is when the path in moduli space that we have taken winds many times around a small periodic circle (without monodromy). In this case, the exact shift symmetry of the axion ensures that nothing dramatic can occur. However, we emphasize that inflation requires a motion in moduli space which is not periodic, either due to monodromy or because the size of the circle is large. In this case, the shift symmetry does not help.

This suggests an alternate perspective on the problem. The periodicity of $\theta$ arises because we can do a large gauge transformation $A_M \to A_M - \partial_M \chi$ for which $\chi$ is not single-valued on the circle but $e^{i \chi}$ is. In particular, we can identify $\theta = 2\pi$ with $\theta = 0$ by performing the transformation
\beq
A_5 \to A_5 - 1/R,~~\chi = y/R,~~\Psi \to e^{-iqy/R} \Psi.
\eeq
While this appears at first glance to be a completely innocent operation, notice that if we are working within an effective field theory with UV cutoff $\Lambda$, this large gauge transformation can bring in modes that are outside the validity of our effective field theory. In particular, if we do not require
\beq \label{eqn:periodicitycondition}
\frac{q}{R} \simlt \Lambda,
\eeq
then the low-frequency modes of the gauge-transformed $\Psi$ field involve very high-frequency modes of the original field, and vice versa. If we do not require~(\ref{eqn:periodicitycondition}), then even the periodicity of $\theta$ becomes a subtle question in the low-energy theory!

Heuristically, another way to see a problem with these large field ranges is to consider the effective mass for $\Psi$~(\ref{eq:effectivemassterm}):
\beq
\int d^5 x \sqrt{-g} \frac{q\theta}{2\pi R} {\bar \Psi} \Gamma^5 \Psi \,.
\eeq
If~(\ref{eq:monodromyfieldrange}) is violated then $\Psi$ receives an effective mass which removes it from the low-energy effective field theory. Of course, the full term involves $\partial_5 - i qA_5$, so the large mass obtained from the Wilson loop can be compensated by high-frequency oscillations in $y$, but these high-frequency modes are not part of the EFT that we started with at the origin of moduli space. We elaborate on this point in appendix \ref{sec:lattice}, using a manifestly gauge-invariant lattice regulator to explore how physical quantities can depend on the cutoff if $\Lambda R$ is not large compared to $q$.
Large effective masses far out on the moduli space are particularly suspect in cases where $\Psi$ plays an important dynamical role. For instance, if $\Psi$ provides one of the dominant instanton contributions to the potential, what does it mean to compute $V(\theta)$ for a value of $\theta$ for which it is inconsistent to keep track of the particle generating the potential? If $\Psi$ is a field that is necessary to satisfy the electric WGC, decoupling it from the effective theory is inconsistent with the EWGC.

We propose a new constraint on theories of extranatural inflation based on this consistency requirement. One statement of the constraint is the following:\\

\noindent
{\bf Single-EFT Consistency Criterion (SECC):} in order to have a controlled description of a portion of the moduli space within a single effective field theory, we demand that any field which is part of the EFT at one point of the moduli space is not decoupled by terms like (\ref{eq:effectivemassterm}) in a different region of the moduli space. Equivalently, if a Kaluza-Klein mode is light somewhere in the moduli space, this mode should exist within the effective theory at the origin of moduli space. This constrains $R \Lambda$ to satisfy (\ref{eq:monodromyfieldrange}).\\

\noindent
Loosely, in a controlled theory a mode cannot appear ``out of the blue.'' This seems to us to be a sufficiently well-motivated criterion that it is worthwhile to explore its consequences. An equivalent statement, if we want to describe the {\em entire} moduli space in a single EFT, is:\\

\noindent
{\bf Single-EFT Axion Periodicity Criterion}: The periodic identification of 4d axions arising from an underlying higher-dimensional gauge theory should arise from large gauge transformations that are well-defined within the higher-dimensional EFT. Specifically, if the theory has a UV cutoff $\Lambda$, then fields which are smooth on scales much larger than $\Lambda^{-1}$ should not oscillate on length scales shorter than $\Lambda^{-1}$ after the gauge transformation.\\

The SECC assumes that we should be able to work with a single well-defined 5d effective field theory. One might imagine a patchwork of effective field theories, each valid over a limited range of $\theta$, which are matched onto each other in overlapping regimes. Nothing intrinsically seems to prevent us from considering the 5d theory on a Wilson line background with any particular value of $\theta$; what we have seen is that connecting the theories at different values of $\theta$ may be difficult. One might consider the case of Seiberg-Witten theory \cite{Seiberg:1994rs}, in which vacua with weakly coupled electrons and with weakly coupled monopoles cannot coexist in the same EFT from the IR point of view but are guaranteed to be smoothly joined together due to well-understood UV physics. Our claim is that because our 5d theory came with a built-in cutoff at $\Lambda$, we do not actually have such a sharp understanding of the UV physics in this case. It may exist if we embed the 5d theory in a more complete UV setting.

If the large gauge transformations that guarantee an identification $\theta \sim \theta + 2\pi$ in the four-dimensional effective field theory are not actually valid operations in the UV completion that we started with, this suggests that we do not truly have a controlled theory of axions. In such a case it is unclear what a computation of the axion potential as a periodic function of the $\theta$'s even means. Nonetheless, we cannot give any fully rigorous argument in favor of the Single-EFT Consistency Criterion. In this paper, we will explore the consequences of the SECC, while welcoming debate on its merits.

\subsection{Consequences of Single-EFT Consistency for monodromy}\label{sec:monodromy}

The Single-EFT Consistency Criterion, in the form of the bound (\ref{eq:monodromyfieldrange}), is a significant potential obstacle to any model based on axion monodromy. To see why, consider any model in which an axion field winds around the circle $N$ times in the presence of monodromy. We have the constraint
\beq
N \simlt R\Lambda
\eeq
from the SECC. But we also have, from the magnetic form of the WGC, the additional constraint
\beq
\Lambda \simlt e M_{\rm Pl}.
\eeq
These conditions together with $f = 1/(2 \pi R\, e)$ imply
\beq
N f \simlt M_{\rm Pl},
\eeq
so the effective {\em total} field range from winding $N$ times around the circle is {\em still} bounded above by the Planck scale. 

Monodromy was important in this argument. Without monodromy, if the physical state were {\em exactly} the same after each trip around the circle, we could get away with only requiring that a gauge transformation $\theta \to \theta + 2\pi$ is well-defined (and then repeat it $N$ times) rather than that a larger field range $\Delta \theta \sim 2\pi N$ is accessible within the effective theory.

Although we have phrased our argument in terms of 1-form gauge fields in five dimensions, a similar constraint will arise from the SECC for the more general $p$-form models. Just as our charged field $\Psi$ obtained an effective Lorentz-violating mass $\sim A_5 {\bar \Psi} \Gamma^5 \Psi$ in the presence of a background gauge field, the presence of a background $p$-form will add a Lorentz-violating tension term to the worldvolume effective theory of a $(p-1)$-brane, potentially decoupling it from the effective field theory.

The SECC argument against monodromy is not airtight. In section \ref{sec:decayconstantalignment} we will consider a two-axion model of inflation in which one axion winds $N$ times around the circle, but we will see in section \ref{subsec:fundamentaldomain} that this does not necessarily violate the SECC. The reason is that there is a compensating contribution to the mass of the charged fields coming from the second axion. This gives some insight into how a monodromy model might successfully escape the SECC. However, in the model we will discuss, the existence of any charged fields with {\em different} charges than those producing the dominant instanton effects will restore the power of the SECC, whereas the one case that evades the SECC is constrained by a different requirement that we will formulate in section \ref{sec:kkimplications}.

\subsection{The Weak Gravity Conjecture across moduli space}\label{sec:kkimplications}

There is one other conjecture, in a similar spirit to the SECC but differing in its details and its implications, that is worth considering:\\

\noindent
{\bf Extended Weak Gravity Conjecture (XWGC):} The weak gravity conjecture should be satisfied at any stationary point of the potential.\\

In fact, we will only use this condition applied at extrema of the potential, rather than generic stationary points. We could also have imposed a stronger condition that the WGC holds {\em everywhere} in the moduli space, though at least for the cases we consider the results would be equivalent. 

This provides a new viewpoint on the small-action loophole. A charged particle with 5d mass $m_5 = 0$ obviously satisfies the (electric) WGC in five dimensions, and we saw that such a particle can generate a potential compatible with inflation despite the lack of exponential suppression of higher harmonics. However, the XWGC demands that the electric WGC is satisfied also at stationary points of the potential away from $\theta = 0$. These are classically stable states, but those that are not local minima will eventually tunnel away from the critical point. Because tunneling can be a slow process, and charged black holes discharge quickly when the WGC is satisfied \cite{Gibbons:1975kk,Schumacher:1985zz,Khriplovich:1999gm}, it seems plausible that the WGC should hold even in these unstable states. As mentioned in the introduction, we suspect that the requirement that black holes decay is actually a dynamical requirement that they shed charge often enough relative to uncharged Hawking quanta, rather than a simple kinematic statement that they can decay at all. Further work on black hole thermodynamics may help to justify or refute the XWGC by quantifying the timescale on which we require charge to be lost.

The Kaluza--Klein modes have masses spaced by $1/R$, so at the maximum of the potential $\theta = \pi/q$ the masses are maximally shifted and the lightest electrically charged particle has $m = 1/(2 R)$, or larger if we begin with $m_5 \neq 0$.
Let us assume,  as in~\S\ref{subsec:elecWGCinflation}, that the same particle which generates the leading contribution to the axion potential satisfies the XWGC. Let us first give a simple heuristic argument for why the XWGC could close the small-action loophole. For a particle of charge $q$, we obtain the bound:
\be
\frac{1}{2 R} < \sqrt{2}q e M_{\rm Pl} = \frac{q M_{\rm Pl}}{\sqrt{2} \pi R f}~~\Rightarrow~~\frac{f}{q} < \frac{\sqrt{2}M_{\rm Pl}}{\pi}.
\ee
Assuming that the same particle generates the axion potential, $f/q$ is precisely the effective axion field range, and the small action loophole appears to have been closed without invoking the magnetic form of the conjecture.\footnote{The argument based on the magnetic WGC is still somewhat stronger, because we don't need to assume that the charged particle which generates the leading contribution to the axion potential has any other special role to play.}

The argument we have just given ignores an important effect. The compactification on the circle produces a second $U(1)$ gauge field, namely the KK $U(1)$ arising from graviton modes with one leg on the circle. At nonzero values of $\theta$, the two $U(1)$ gauge fields mix and the correct WGC to consider is the convex hull condition applied to our original $U(1)$ gauge theory and the Kaluza-Klein $U(1)$. We present a detailed derivation of this statement and discussion of the mixing effect in \cite{Heidenreich:2015nta}. The weak gravity bound becomes
\be
m^2 \leq \gamma e^2 q^2 M_{\rm Pl}^2 + \frac{g_{\rm KK}}{R^2} \left(n - \frac{q \theta}{2\pi}\right)^2.
\ee
The constant $\gamma$ is $2$ as above if the radion mode is stabilized, but is $3/2$ if the radion is unstabilized. Similarly, the constant $g_{\rm KK}$ is $1$ for an unstabilized radion and $2$ for a stabilized radion. If a 5d particle obeys the WGC, then any of its KK modes in 4d will in fact obey this inequality for any value of $\theta$, undermining the heuristic argument we gave above. However, our conclusion survives once we take the convex hull condition into account.

The reason we obtain a bound from the convex hull condition is that there is not a KK mode in every direction in charge space. We want to apply the convex hull condition to the charge-to-mass vectors
\begin{align}
{\vec z} = \left(z, z_{\rm KK}\right) &= \frac{1}{m(n,\theta)} \left(\sqrt{\gamma} e q M_{\rm Pl},~\frac{g_{\rm KK}}{R} \left(n - q \btheta\right)\right), \\
m(n,\theta) &= \sqrt{m_5^2 + \frac{1}{R^2}\left(n - q\btheta\right)^2}.
\end{align}
Without loss of generality, we specialize to the case $q = 1$ for simplicity and set $f = 1/(2 \pi e R)$. We choose $\btheta = 1/2$ so that the KK charge of the particles is $n - \btheta$, an odd half-integer. We take $m_5 = 0$ to study the small-action loophole. Any other value of $m_5$ will, for fixed charges $q$ and $n$, lead to a shorter vector ${\vec z}$ and thus a tighter constraint. We have a set of charge-to-mass vectors 
\be
{\vec z}_n = \left(\frac{1}{n-\frac{1}{2}} \frac{\sqrt{\gamma} M_{\rm Pl}}{2 \pi f},~ {\rm sgn}\left(n-\frac{1}{2}\right)\, g_{\rm KK}\right).
\ee
We also have a set of charge-to-mass vectors from KK gravitons or dilatons, which are uncharged under the $U(1)$ but carry KK charge, and for unstabilized radion will always saturate the WGC for their direction in the charge lattice:
\be
{\vec z}_{{\rm grav};n} = \left(0,~{\rm sgn}(n)\, g_{\rm KK}\right).
\ee
All of these vectors are outside the open unit ball, so in the direction of any ${\vec z}_n$, we satisfy WGC. But of course the striking thing about these vectors is that they all have a ``$\pm g_{\rm KK}$'' in the second entry. That is, the KK charge always satisfies the WGC bound (saturating it when the radion is unstabilized), and we're at a point on the moduli space where {\em every} charged particle has KK charge due to the axion effect.

\begin{figure}[!h]
\begin{center}
\scalebox{1.4}{
\begin{tikzpicture}[line width=1.5 pt,
   axis/.style={very thick, ->, >=stealth'}]
\draw[axis] (-1.5,0)--(1.5,0) node(xline)[right] {$z$};
\draw[axis] (0,-1.5)--(0,1.5) node(xline)[right] {$z_{\rm KK}$};
\draw[dashed] (0,0) circle (1.0);
\filldraw[orange] (0,1.0) circle (0.04);
\filldraw[orange] (0,-1.0) circle (0.04);
\filldraw[blue] (1.1-0.015,1.0-0.015) rectangle (1.1+0.015,1.0+0.015);
\filldraw[blue] (-1.1-0.015,1.0-0.015) rectangle (-1.1+0.015,1.0+0.015);
\filldraw[blue] (1.1-0.015,-1.0-0.015) rectangle (1.1+0.015,-1.0+0.015);
\filldraw[blue] (-1.1-0.015,-1.0-0.015) rectangle (-1.1+0.015,-1.0+0.015);
\filldraw[blue] (0.367-0.015,1.0-0.015) rectangle (0.367+0.015,1.0+0.015);
\filldraw[blue] (-0.367-0.015,1.0-0.015) rectangle (-0.367+0.015,1.0+0.015);
\filldraw[blue] (0.367-0.015,-1.0-0.015) rectangle (0.367+0.015,-1.0+0.015);
\filldraw[blue] (-0.367-0.015,-1.0-0.015) rectangle (-0.367+0.015,-1.0+0.015);
\filldraw[blue] (0.22-0.015,1.0-0.015) rectangle (0.22+0.015,1.0+0.015);
\filldraw[blue] (-0.22-0.015,1.0-0.015) rectangle (-0.22+0.015,1.0+0.015);
\filldraw[blue] (0.22-0.015,-1.0-0.015) rectangle (0.22+0.015,-1.0+0.015);
\filldraw[blue] (-0.22-0.015,-1.0-0.015) rectangle (-0.22+0.015,-1.0+0.015);
\filldraw[blue] (0.157-0.015,1.0-0.015) rectangle (0.157+0.015,1.0+0.015);
\filldraw[blue] (-0.157-0.015,1.0-0.015) rectangle (-0.157+0.015,1.0+0.015);
\filldraw[blue] (0.157-0.015,-1.0-0.015) rectangle (0.157+0.015,-1.0+0.015);
\filldraw[blue] (-0.157-0.015,-1.0-0.015) rectangle (-0.157+0.015,-1.0+0.015);
\node at (1.5,1.2) {\tiny $(z_1, 1)$};
\end{tikzpicture}
}
\end{center}
\caption{How the XWGC closes the small-action loophole when $\theta = \pi$. We depict the case $g_{\rm KK} = 1$ for convenience. The horizontal axis is the charge-to-mass ratio $z$ for the $U(1)$ gauge group giving rise to the axion. The vertical axis is the charge-to-mass ratio for Kaluza-Klein charge. The points on the vertical axis (orange circles) correspond to graviton KK modes. The points off the axis (blue squares) correspond to charged particle KK modes, which as $n \to \infty$ accumulate near the orange points. We see that the convex hull condition demands that the horizontal coordinate $z_1$ at $n = 1$ be $\geq 1$, leading to the bound $f < \frac{\sqrt{\gamma}}{\pi} M_{\rm Pl}$.}
\label{fig:xwgcsmallaction}
\end{figure}
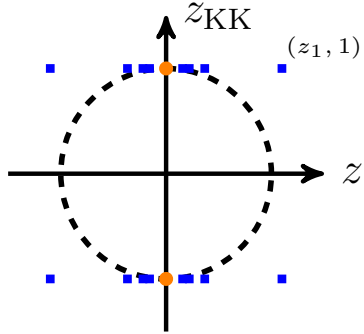

The convex hull requirement for these vectors is depicted in figure \ref{fig:xwgcsmallaction}. We see that the requirement that the unit circle is contained in the convex hull imposes $z_1 \geq 1$, i.e.
\be
f < \frac{\sqrt{\gamma}}{\pi} M_{\rm Pl}.
\ee
If this requirement is not satisfied, a black hole with charge vector $(Q, 0)$ under the two $U(1)$ gauge groups will not be able to decay. Notice that we can easily construct such charge vectors from combinations of KK modes with opposite signs of the KK number. Thus, we see that the XWGC (in its convex hull formulation) closes the small-action loophole for a single axion, despite the subtleties introduced by the KK $U(1)$.

It is clear from figure \ref{fig:xwgcsmallaction} that, because the value of $z_{\rm KK}$ is the same for every state, we can simply project the problem down to the lower-dimensional problem of considering only $z$. This will continue to be true in a scenario where we consider multiple $U(1)$ gauge bosons: a theory of $n$ gauge bosons in 5d gives rise to an ($n+1$)-dimensional convex hull problem in 4 dimensions, but so long as we go to a point on the moduli space where every charged particle in the theory has a common value (up to a sign) for $z_{\rm KK}$, the added KK direction can be projected out of the argument. We will exploit this below in discussing a scenario with two $U(1)$ groups in 5d.

Our main motivation for studying the SECC and XWGC is to apply them to a version of alignment inflation which the usual WGC is not strong enough to exclude. We will consider this scenario in the next section.

\subsection{The Lattice Weak Gravity Conjecture}\label{sec:LWGC}

In \cite{Heidenreich:2015nta}, we introduced another form of the WGC:\\

\noindent
{\bf The Lattice Weak Gravity Conjecture (LWGC)}: At every point in the charge lattice, there exists a state that satisfies the WGC.\\

\noindent
This conjecture is obeyed, for instance, by string states of the $SO(32)$ heterotic string as well as Kaluza-Klein reduction of pure gravity on a torus.

At first, this conjecture seems to highlight an apparent loophole in our arguments against $N$-flation in section \ref{sec:multiaxion}. If there is an instanton for every point on the charge lattice lying outside the unit ball, then the leading instantons can lie on or just outside the unit ball, implying 
actions which are larger by a factor of $\sqrt{N}$ relative to the case where only $N$ instantons satisfy the convex hull condition.

However, the large number of subleading instantons required by the LWGC are sufficient to spoil the flatness of the potential.
In particular, consider a theory with $N$ axions that just marginally satisfies the LWGC in every possible direction.  This implies the existence of infinite tower of instantons of every possible integral charge $\vec{Q} = (Q_1,...,Q_N)$.  Further, setting the actions of the leading instantons to be $\mathcal{O}(1)$, and setting each of the decay constants to be Planckian, $f_i \sim M_{\rm Pl}$, we have
\begin{equation}
f_{\rm eff} \sim \sqrt{N} M_{\rm Pl}.
\end{equation}
The convex hull condition is satisfied because of the infinite number of subleading instantons, which densely fill the unit sphere.  A necessary condition for this inflationary model is that the subleading instanton actions must scale with the instanton charges as
\begin{equation}
S_{\vec{Q}}^2 \sim \sum_i Q_i^2.
\label{quadrature}
\end{equation}
If, on the other hand, the actions were to grow linearly with the charges, $S_{\vec{Q}} \sim \sum_i |Q_i|$, then the subleading instantons would not densely fill in the unit sphere but instead a cube of diagonal length $M_{\rm Pl}$ centered at the origin.  This would not contain the unit ball, so the convex hull condition would be violated.

However, the particular growth of the instanton actions in (\ref{quadrature}) that allows this scenario to satisfy the LWGC is also what leads to its downfall.  The instanton actions are smaller in this scenario than in a model in which the instanton actions grow linearly, which means their contributions to the potential are larger.  We will now show that this enhancement yields large corrections to the inflationary potential, making it unsuitable for inflation.

For this, we need only consider the special class of instantons with charges $(\pm 1, ... \pm 1)$.  Of course, there are many more instantons that will contribute to the inflationary potential, but for our purposes it suffices to show that just the instantons in this special class combine to give large contributions.  There are $2^N$ such instantons, which have action $S \approx \sqrt{N} S_{\rm leading}$ relative to the actions of the leading instantons, $S_{\rm leading}$.  The $N$-flation potential generated by the leading instantons is of the form
\begin{equation}
V(\phi_i) \supset \mathcal{A} e^{-S_{\rm leading}} \cos (\phi_i/f)
\label{eq:leadingpotential}
\end{equation}
The subleading instantons under consideration, on the other hand, give a potential of the form
\begin{equation}
V(\phi_i) \supset \mathcal{A} e^{-\sqrt{N} S_{\rm leading}} \cos (\sum_i \eta_i \phi_i/f).
\end{equation}
Here $\eta_i = \pm 1$ depending on which instanton is being considered.  We consider inflation along the diagonal direction $\phi_i = \phi$.  A necessary condition for inflation is that the potential contributions from these instantons must in fact be negligible compared to those from the leading instantons.  However, there are $2^N$ instantons of the form in question, and each one introduces a potential contribution of magnitude $e^{-\sqrt{N} S_{\rm leading}}$.  Thus, the total potential contribution from these instantons grows roughly as $2^N e^{-\sqrt{N} S_{\rm leading}}$.  To protect the $N$-flation potential of (\ref{eq:leadingpotential}) from these subleading effects at large $N$, we must therefore take $S_{\rm leading} \gtrsim \sqrt{N}$.  However, the LWGC implies $f \lesssim M_{\rm Pl}/S_{\rm leading}$, so $f_{\rm eff} \sim \sqrt{N} f \lesssim M_{\rm Pl}$.  Hence, parametric enhancement of the effective decay constant via isotropic $N$-flation is inconsistent with the LWGC.  The LWGC also restricts models of decay constant alignment, which we discuss in the following section.

\section{Decay constant alignment}
\label{sec:decayconstantalignment}

\begin{figure}[!h]
\begin{center}
\scalebox{0.85}{
\begin{tikzpicture}[line width=1.5 pt,
   axis/.style={very thick, ->, >=stealth'}]
\draw[fill=blue, opacity=0.2] (0,-0.75)--(0,0.75)--(-1.5,8.25)--(-1.5,6.75)--cycle;
\draw[axis] (-3,0)--(3,0) node(xline)[right] {$\theta_A$};
\draw[axis] (0,-0.75)--(0,8.0) node(xline)[right] {$\theta_B$};
\draw[blue,very thick, ->, >=stealth'] (-0.75,3.75)--(0,0);
\node at (1.5,0) {|};
\node at (1.5,-0.5) {$2\pi$};
\node at (-1.5,0) {|};
\node at (-1.5,-0.5) {$-2\pi$};
\node at (0,1.5) {$-$};
\node at (0.5,1.5) {$2\pi$};
\node at (0,3.0) {$-$};
\node at (0.5,3.0) {$4\pi$};
\node at (0,4.5) {$-$};
\node at (0.5,4.5) {$6\pi$};
\node at (0,6.0) {$-$};
\node at (0.5,6.0) {$8\pi$};
\node at (0,7.5) {$-$};
\node at (0.5,7.5) {$10\pi$};
\node[red] at (-1.05,4.5) {$\circ$};
\node[red] at (-1.05,6.0) {$\circ$};
\node[purple] at (0.0,-0.5) {$\times$};
\node[purple] at (-1.5,7.0) {$\times$};
\begin{scope}[shift={(6.5,1.0)}]
\draw[fill=blue, opacity=0.2] (0,0)--(3,0)--(3,3)--(0,3)--cycle;
\draw[blue,very thick, ->, >=stealth'] (2.4,3)--(3,0);
\draw[blue,very thick, ->, >=stealth'] (1.8,3)--(2.4,0);
\draw[blue,very thick, ->, >=stealth'] (1.5,1.5)--(1.8,0);
\draw[axis] (-1,0)--(5,0) node(xline)[right] {$\theta_A$};
\draw[axis] (0,-1)--(0,5) node(xline)[right] {$\theta_B$};
\node at (3,0) {|};
\node at (3,-0.5) {$2\pi$};
\node at (0,3.0) {$-$};
\node at (-0.5,3.0) {$2\pi$};
\end{scope}
\end{tikzpicture}
}
\end{center}
\caption{Two views of the fundamental domain (shaded) of the two axions for the case $N = 5$, together with a trajectory (thick blue arrow) beginning at a maximum of the potential and ending at the origin. For clarity, the view on the left and right are not drawn to the same scale. The right hand view is a ``natural'' parametrization with $0 \leq \theta_{A,B} \leq 2\pi$ but requires that we discontinuously change the value of $\theta$ and execute a corresponding monodromy on the Kaluza--Klein spectrum when wrapping around the torus. The left-hand view chooses a parametrization in which the values of $\theta_{A,B}$ and the Kaluza-Klein masses change smoothly during all of inflation. To illustrate the periodic identifications imposed on the boundaries, we show two points labeled with a red $\color{red} \circ$ that are identified and two points labeled with a purple $\color{purple} \times$ that are identified.}
\label{fig:axiondomain}
\end{figure}
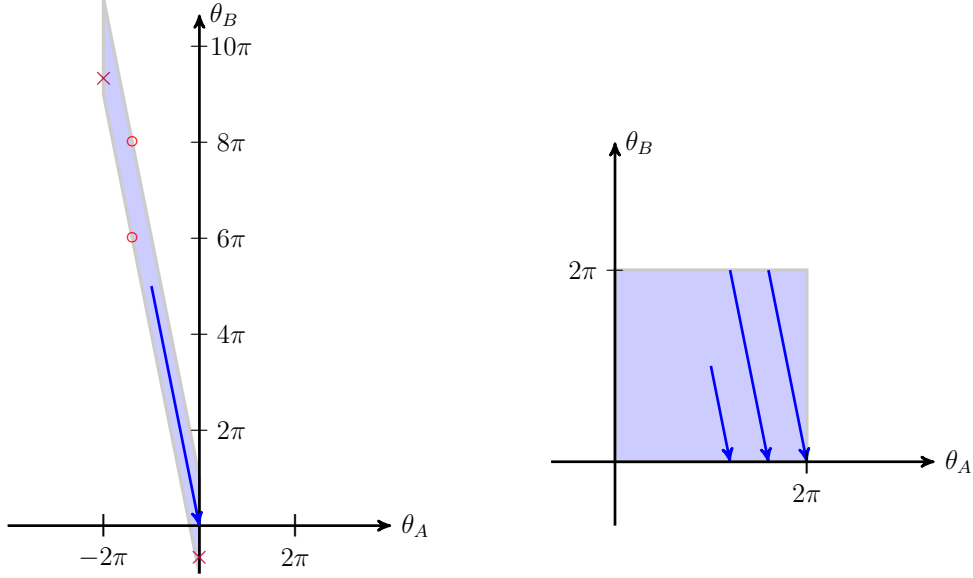

The arguments we have given in section \ref{sec:multiaxion} break down when the charge assignments of the dominant instanton effects are not small integers in the same basis where the magnetic monopole charges satisfying the magnetic WGC are. This case is especially subtle. For concreteness, we will focus the two-axion model of \cite{delaFuente:2014aca}. We assume a basis in which two compact axion fields, $\theta_A$ and $\theta_B$, have diagonal kinetic terms with decay constants $f_A$ and $f_B$:
\be
{\cal L}_{\rm kin} = \frac{1}{2} f_A^2 \partial_\mu \theta_A \partial^\mu \theta_A + \frac{1}{2} f_B^2 \partial_\mu \theta_B \partial^\mu \theta_B.
\ee
We further assume that in this basis the magnetic monopoles satisfying WGC have charge assignments $(1,0)$ and $(0,1)$, leading to the constraints
\beq
f_A, f_B \simlt M_{\rm Pl}.
\eeq
However, we assume that the dominant instanton effects arise from electric charges $(1,0)$ and $(N,1)$ that are highly aligned in this basis. As a result, the potential behave as
\beq
V(\theta_A, \theta_B) \approx V_0 \left(1-\cos(\theta_A)\right) + {\tilde V}_0 \left(1 - \cos(N \theta_A + \theta_B)\right).
\eeq
As emphasized in \cite{delaFuente:2014aca}, this provides a UV setting for the alignment mechanism of Kim, Nilles, and Peloso \cite{Kim:2004rp} which appears to satisfy the WGC constraint. Inflation occurs on a trajectory for which $\theta_A$ winds once around the circle while $\theta_B$ winds $N$ times, leading to an effective decay constant
\beq
f_{\rm eff} \approx N f_B.
\eeq 
This inflationary trajectory is illustrated in figure \ref{fig:axiondomain}. The instantons can be generated from worldlines of charged particles, but \cite{delaFuente:2014aca} also discusses a scenario in which the factor of $N$ can be the level of a Chern-Simons coupling $A \wedge G^a \wedge G^a$ in the 5d theory, potentially arising from a quantized flux in an even higher-dimensional theory.

\subsection{Parametrizing the fundamental domain}
\label{subsec:fundamentaldomain}

If we consider the axion fields to range over $0 \leq \theta_A, \theta_B \leq 2\pi$, as depicted in the right-hand side of figure \ref{fig:axiondomain}, then the periodic identification $\theta_A \to \theta_A + 2\pi$ that wraps the left-hand side of the square onto the right-hand side shifts the mass of a state with charge $(N,1)$ by $N/R$, suggesting the possibility that a cutoff near $1/R$ is too low for consistency. On the other hand, the inflaton trajectory winds multiple times, and we can think of this large mass shift as a consequence of the monodromies induced every time we wrap around the torus and shift our coordinate $\theta$ discontinuously.

A more useful parametrization of the moduli space is to ``unwind'' it so that it is aligned with the inflaton trajectory, as in the left panel of figure \ref{fig:axiondomain}. In this case, moving off the right edge of the space wraps back to the left edge at a lower point, corresponding to $\theta_B \to \theta_B - 2\pi$ with $\theta_A$ unchanged. The identification of the upper left edge with the bottom right edge corresponds to $(-2\pi,\theta_B) \to (0, \theta_B - 2 N \pi)$. This is a large change in $\theta_B$, but it leaves $N\theta_A + \theta_B$ fixed, so for a particle of charge $(N,1)$ there is no change in the mass spectrum for this transformation. Thus, if the only particles in our effective theory have charges $(1,0)$ and $(N,1)$, the SECC imposes no obstacle to taking $\Lambda \sim 1/R$. The masses of the Kaluza--Klein modes of these fields are at most of order $1/R$ throughout the moduli space.

This illustrates a general fact: the physical criterion that we would like to impose is that there is a single effective theory that is valid everywhere on the moduli space. Some parametrizations of the moduli space might obscure the existence of this theory, while others make it manifest.

\subsection{Case 1: dominant instantons satisfy electric WGC}\label{subsec:case1xwgc}

The first case we consider is that the same charges $(1,0)$ and $(N,1)$ that control the axion potential also are responsible for satisfying the electric WGC. The argument of \cite{delaFuente:2014aca}, emphasized to us by the authors \cite{Prashant}, is that once the magnetic WGC constraint that $f_A, f_B \simlt M_{\rm Pl}$ is imposed, there is no further WGC constraint. The charged particles can be arbitrarily light, and direct calculation confirms that higher instanton contributions are numerically small. However, this changes if we impose our stronger XWGC conjecture from section \ref{sec:kkimplications}. The potential has a local maximum where the arguments of both cosines are $\pi$, i.e. $N \theta_A + \theta_B = \pi$ and $\theta_A = \pi$. At this point, the lightest KK modes for both charged particles have mass $1/(2 R)$ (assuming that $m_5 = 0$). As in the XWGC discussion above, there is also a third, Kaluza-Klein, $U(1)$ gauge group to consider, but at this point in moduli space every charged particle in the theory has $z_{\rm KK} = \pm g_{\rm KK}$, so the third dimension can be projected out of the argument. Thus the XWGC tells us that, assuming we started with the best-case scenario where the charged particles have negligible 5d masses, the charge-to-mass vectors at the maximum of the potential (and indeed at any generic point in moduli space) are of order
\beq
{\vec z}_1 & = & \left(\frac{\sqrt{2} e_A M_{\rm Pl}}{m_1}, 0\right) \sim \left(e_A R M_{\rm Pl}, 0\right) \sim \left(\frac{M_{\rm Pl}}{f_A}, 0\right), \nonumber \\
{\vec z}_2 & = & \left(\frac{\sqrt{2}  N e_A M_{\rm Pl}}{m_2}, \frac{\sqrt{2}  e_B M_{\rm Pl}}{m_2}\right) \sim \left(\frac{N M_{\rm Pl}}{f_A}, \frac{M_{\rm Pl}}{f_B}\right).
\eeq
Now we require that the convex hull of the vectors $\pm {\vec z}_1, \pm {\vec z}_2$ contains the unit sphere. The line passing through the points $(-\alpha, 0)$ and $(N\alpha, \beta)$ in the $(x,y)$ plane is $(N+1) \alpha y = \beta (\alpha + x)$, so if we read off where this crosses the $y$-axis, we obtain the constraint $\beta \geq N+1$. Substituting the vectors we're interested in,
\beq
\frac{M_{\rm Pl}}{f_B} \simgt N+1~~\Rightarrow~~f_{\rm eff} = N f_B \simlt M_{\rm Pl}.
\eeq
This shows that our conjecture that the electric WGC should be satisfied at all stationary points in the axion moduli space is strong enough to exclude decay constant alignment in the scenario where the same particles are responsible for both satisfying the WGC and supplying the dominant instanton effects.

The LWGC postulates the existence of an instanton satisfying the WGC at every point on the charge lattice.  If this version of the WGC is true, we must therefore consider models with additional instantons.  We now turn our attention to these models.

\subsection{Case 2: additional particles satisfy electric WGC}\label{subsec:case2secc}

On the other hand, we could consider a different scenario. The reason we obtained such a large bound on $f_B$ from the convex hull condition is that its charge-to-mass vector was nearly aligned with that of $f_A$. This was necessary to obtain a large field range in inflation, but what if we satisfy the convex hull condition (and hence the electric WGC) with {\em different} vectors than the ones that dominate the instanton contributions? This possibility was, again, suggested to us by the authors of \cite{delaFuente:2014aca}. Suppose that we have three relevant charges, $(1,0), (N,1),$ and $(0,1)$. The first two supply the dominant instanton contributions, while the first and third satisfy the WGC. We can take the 5d mass of the $(0,1)$ particle to be somewhat large compared to $1/R$, so that its instanton contribution is suppressed, but not parametrically large by a factor of $N$, so that we do not shrink its mass-to-charge vector by enough to obtain the desired bound on $f_B$.

This is the point at which the SECC becomes crucial. We can consider which modes are present in our effective theory as we move around the moduli space. The particles with charge $(q_A, q_B)$ have their masses shifted by $\frac{1}{2\pi R}\left(q_1 \theta_A + q_2 \theta_B\right)$. As discussed in section \ref{subsec:fundamentaldomain}, in order to adiabatically track the mass of a particular mode, we should work in the ``unwound'' moduli space where the inflaton trajectory is continuous, as in the left-hand panel of figure \ref{fig:axiondomain}. In this fundamental domain neither $\theta_A$ nor $N \theta_A + \theta_B$ is parametrically large, but $\theta_B$ itself is. Precisely because the inflaton direction winds around the $\theta_B$ circle multiple times, when we unwind the moduli space we find that $0 \leq \theta_B \leq 2\pi N$. Thus, the new particle of charge $(0,1)$ that we invoked to satisfy the electric WGC without running into difficulty with the convex hull condition is not part of a single consistent effective field theory defined over the entire axion moduli space unless we satisfy the bound
\beq
2\pi R \Lambda \simgt 2 \pi N.
\eeq
Again, this is the precise parametric bound that we need to obtain $f_{\rm eff} \simlt M_{\rm Pl}$. In fact, in this particular case we do not even have to invoke the full SECC. We only need to require a consistent set of modes {\em along the inflaton trajectory}. This leaves open the possibility that this weaker requirement and the XWGC are sufficient assumptions, without the full SECC. 

Notice that this argument, unlike many of our previous arguments, relies on the EWGC: if the state of charge $(1,0)$ which satisfies the convex hull condition is never in the effective theory to begin with, then the SECC does not further constrain the model. However, so long as there is \emph{some} effective field theory description of this state, the SECC will rule out the model, whereas theories which violate the EWGC may present other control problems, as explained in the introduction.

\subsection{Status of decay constant alignment}

To summarize, we have excluded the particular example of decay constant alignment with charges $(1,0)$ and $(N,1)$ using the two conjectured constraints from section \ref{sec:kkmonodromy}. The XWGC is crucial for avoiding the small-action loophole in which we use the same light fields to generate the dominant instanton effects and to satisfy the electric WGC. The SECC is needed for the case when the particles generating the instanton effects and satisfying the electric WGC are not the same. In both cases, the parametric constraint that we extract is {\em precisely} what is necessary to obtain a maximum field range of $M_{\rm Pl}$. This is, at the least, very suggestive. It calls for further effort to understand how strongly motivated our conjectured constraints are. Again, the possible counterargument to the XWGC is that stationary points in the moduli space that are not minima have a finite lifetime before tunneling to a lower point on the potential, so arguments based on concerns about exactly stable remnants are not directly relevant. For the SECC, the concern is that despite the inconsistency of a single effective field theory, an ultraviolet completion may somehow allow a patchwork theory to be constructed. We do not have definitive rebuttals to these possible counterarguments, but our constraints appear plausible and well-motivated to us, and it is very interesting that the XWGC and the SECC precisely exclude a model that otherwise appears to sail through the WGC's tests with flying colors.

Throughout this section we have referred to charged particles of charge $(N,1)$. As explained in \cite{delaFuente:2014aca}, it may be possible to generate instanton effects through other means, such as Chern-Simons couplings to nonabelian gauge theories that confine. This will not affect our arguments. Some set of charged particles must exist to satisfy the electric WGC, and even if they do not contribute significant instanton effects, their charge assignments still lead to uncontrolled large gauge transformations at the boundaries of moduli space.

We have not attempted to derive a bound in detail for arbitrary charge assignments or arbitrary numbers of axions, but by putting together the ideas of this section with those we employed in section \ref{sec:multiaxion}, we expect that any model based on compact axion fields with no crucially new physics idea as input can be excluded. 

\section{Do Entropy Bounds Exclude Large-Field Inflation?}
\label{sec:entropybounds}

One constraint widely believed to apply in any theory of quantum gravity is an entropy bound, loosely speaking that the logarithm of the number of microstates accessible to a system bounded by a surface of area $A$ is at most of order $A M_{\rm Pl}^2$. Such conjectures originated with Bekenstein~\cite{Bekenstein:1980jp} and were given a sharp covariant form by Bousso~\cite{Bousso:1999xy}. Entanglement entropy in quantum field theory~\cite{Srednicki:1993im,Casini:2009sr} has been shown to satisfy such a bound~\cite{Casini:2008cr,Bousso:2014sda,Bousso:2014uxa}, suggesting that it will be difficult to place theories in the swampland simply by arguing that they violate an entropy bound. On the other hand, several authors have argued for precisely such statements regarding large-field inflation models~\cite{Kaloper:1999tt,Conlon:2012tz,Boubekeur:2013kga}, including recently in the context of the Weak Gravity Conjecture~\cite{Brown:2015iha}. Our goal in this section is to critically review these arguments. We find that they rely on unjustified assumptions. Our assessment is that models of large-field inflation are consistent with entropy bounds.

\subsection{The reheating objection}

An argument against theories with a large number of $e$-folds has been given based on entropy production during reheating~\cite{Kaloper:1999tt,Brown:2015iha}. Essentially, the objection is that to the extent that we can trust the semiclassical picture of the post-reheating universe as a radiation-dominated phase, it will contain a hot plasma with entropy per unit volume $s \propto T^3$. Because the entropy scales with volume and the entropy bound scales with area, it seems there is a potential conflict. But if we consider a radiation-dominated universe,
\beq
\rho \sim H^2 M_{\rm Pl}^2 \sim T^4,
\eeq
so in a volume of radius $R \sim H^{-1}$, the entropy associated with the radiation is
\beq
S \sim T^3 R^3 \sim \frac{1}{T} H^2 M_{\rm Pl}^2 H^{-3} \sim \frac{M_{\rm Pl}^2}{H^2} \frac{H}{T} \sim S_{\rm Bek} \frac{T}{M_{\rm Pl}}.
\eeq
Thus, the Bekenstein bound is safely satisfied in a Hubble-size volume provided the temperature is much less than the Planck energy, which is surely true whenever effective field theory is valid.

The potential to derive a contradiction between inflation with many $e$-folds and entropy bounds arises from considering volumes of radius $R \gg H^{-1}$. \cite{Kaloper:1999tt} framed the problem in terms of the particle horizon, which grows exponentially during inflation, so the ratio of volume to area becomes much larger than for a Hubble-sized volume. The uncertainty about what region to apply entropy bounds to was a major motivating reason for Bousso's covariant entropy conjecture~\cite{Bousso:1999xy}, which had a significant impact because it gave a precise criterion for which such problems do not arise. Given that substantial evidence has accumulated that Bousso's variation of the entropy bound is the correct criterion, it does not appear that reheating after inflation is problematic even in theories with many $e$-folds.

\subsection{Bit-counting arguments}
\label{sec:bitcounting}

Another argument that axions with super-Planckian decay constants can violate entropy bounds arises from a bit-counting argument. We impose an ultraviolet cutoff on the theory by defining a minimum distance scale $\ell$ and count pixels of area $\ell^2$ on the horizon. There are $\sim 1/(\ell^2 H^2)$ such pixels. This exceeds the bound if $\ell < M_{\rm Pl}^{-1}$. If we keep the UV cutoff of our theory below the Planck scale, there should be no problem.

The UV cutoff in refs.~\cite{Conlon:2012tz,Boubekeur:2013kga} was taken to be $\ell \sim f_a^{-1}$. The argument was that the free-particle two-point function of our axion field, $\left<a(x) a(0)\right> \sim \frac{1}{x^2}$, can no longer be a valid estimate at $x \simlt f_a^{-1}$, because it grows without bound even though the field itself is compact and satisfies $a(x) < 2 \pi f_a$. This is a reasonable argument. We could refine it a bit, by noting that the two-point function is not physical because $a(x)$ is not gauge invariant, but $e^{i a(x)/f_a}$ is, so we can build a Lagrangian out of the latter and expand to find non-renormalizable interactions suppressed by the scale $f_a$, which fixes the cutoff. Although the argument that compactness of the field space implies a UV cutoff is correct, it only imposes an {\em inequality}, $\ell > f_a^{-1}$. A contradiction with entropy bounds arises only if we take this to be an equation. But of course, in any theory of quantum gravity we expect it will not be sensible to talk about distances below the Planck length, so we should never take $\ell < M_{\rm Pl}^{-1}$. Nothing in this argument precludes the possibility of theories with a range of scales $f_a \gg M_{\rm Pl} \gg \ell^{-1}$.

\subsection{The classical entropy current argument}

A further argument considered the inflaton as a perfect fluid and constructed the entropy density from its stress energy tensor~\cite{Boubekeur:2013kga}. This leads to an equation
\beq
{\dot S}_\phi = \frac{8\pi^2}{H^3} {\dot \phi}^2.
\eeq
\cite{Boubekeur:2013kga} then gives an argument with the following logical structure: $\Delta S_\phi$ is computed by integrating the above derivative. The expression for $\Delta S_\phi$ is broken into two pieces and bounded by considering the absolute value of these two pieces. This establishes that $\Delta S_\phi < S_{\rm max} \sim f_a^2/H^2$. It is then observed that $S_{\rm max} > S_{\rm Bek}$ for large-field inflation. But, without some estimate of {\em how close} $\Delta S_\phi$ actually comes to $S_{\rm max}$, this proves nothing.

In fact, we can obtain a much better estimate of $\Delta S_\phi$. For slow-roll inflation we have $\dot \phi \approx -\frac{1}{3H} \frac{\partial V}{\partial \phi}$, so we can rewrite the integral as:
\beq
\Delta S_\phi = \int dt \frac{8\pi^2}{H^3} {\dot \phi} \left(-\frac{1}{3H} \frac{\partial V}{\partial \phi}\right) \approx -\frac{8 \pi^2}{3 H^4} \int d\phi \frac{\partial V}{\partial \phi} \approx \frac{8 \pi^2}{3 H^4} \left|\Delta V\right|,
\eeq
with the approximation that $H$ is relatively constant during the time period considered. Given that during inflation $V \approx 3 H^2 M_{\rm Pl}^2$, this is exactly consistent with the Gibbons-Hawking entropy of a Hubble patch of de Sitter space~\cite{Gibbons:1977mu}, so no bound is violated. In fact, a detailed study of de Sitter thermodynamics in the context of inflation, including perturbations, was undertaken some time ago by Frolov and Kofman, who found no inconsistencies~\cite{Frolov:2002va}.

\section{Assessing strong conjectures}
\label{sec:strongforms}

The original weak gravity paper \cite{ArkaniHamed:2006dz} posited a ``strong form'' of the WGC, which stipulates that in theories with a single $U(1)$, the \emph{lightest} charged particle in the spectrum, and not just any charged particle, should have a charge-to-mass ratio larger than that of an extremal black hole.  As pointed out in \cite{delaFuente:2014aca,Prashant} and later discussed in \cite{Rudelius:2015xta, Brown:2015iha, Bachlechner:2015qja, Hebecker:2015rya, Brown:2015lia}, there is a loophole in the electric WGC which leaves open the possibility of axion inflation, but which would be closed by this strong form of the WGC.  In particular, consider a theory with a single axion $a$ and two instantons of action $S_1 \ll S_2$ and associated decay constants $f_1$, $f_2$, respectively.  Each instanton will introduce a term in the axion potential of the form,
\beq
V(a) \supset \mathcal{A}_i e^{-S_i} \cos a/f_i,
\label{Veq}
\eeq
with $\mathcal{A}_i$ some coefficients.  Now, as long as $f_2 S_2 < M_{\rm Pl}$, the ordinary, mild form of the WGC will be satisfied.  $f_1$ is left unbounded, and since $S_1 \ll S_2$, the potential contributions from the first instanton will dominate those from the second.  On the other hand, the strong form of the WGC requires that the instanton of smaller action, which is $S_1$ in our scenario, must also satisfy the bound $f_1 S_1 < M_{\rm Pl}$.  Thus, if we assume $S_1 > 1$, we find that $f_1$ is constrained to be sub-Planckian, and the axion is unsuitable for inflation.

However, there are a few problems with invoking the ``strong form'' of the WGC to close such loopholes.  First off, straightforwardly generalizing the strong form to theories with multiple $U(1)$s proves problematic, as it implies constraints on the spectrum that are clearly far too strong.  To see why, consider a very simple theory with two $U(1)$s and two particles of mass $m_1$, $m_2$ with charges $(q_1,0)$ and $(0,q_2)$.  If one considers either of the $U(1)$'s in this basis, the na\"ive ``strong form of the WGC'' holds that lightest particle charged under each $U(1)$ should have $q/m > 1/M_{\rm Pl}$.  However, suppose we now make a very small basis rotation of our $U(1)$s, so that particles 1 and 2 now have charges
$$(q_1+ \mathcal{O}(\varepsilon^2),q_1 \varepsilon + \mathcal{O}(\varepsilon^2))\,,~~(-q_2 \varepsilon+ \mathcal{O}(\varepsilon^2), q_2 + \mathcal{O}(\varepsilon^2)),$$
respectively.  In this new basis, the statement that the lightest particle charged under each $U(1)$ should have sufficiently large charge-to-mass ratio is problematic: if $m_1 < m_2$, then by taking $\varepsilon$ small enough, we can ensure that $q_1 \varepsilon / m_1$ is too small to satisfy the bound.  If $m_2 < m_1 $, then we can do the same with $q_2 \varepsilon / m_2$.  The only way this conjecture could hold is if $m_1 =m_2$ i.e. if every particle in the spectrum has precisely the same mass.  This is clearly unacceptable.

This problem may be remedied.  Namely, we may define the ``strong form'' of the WGC for theories with $N$ $U(1)$s to be the statement that the lightest particles whose charge-to-mass vectors span the full $\mathbb{R}^N$ should satisfy the convex hull condition.  It is easy to check that in a theory with a single $U(1)$, this definition of the strong WGC reduces to the usual one.  Furthermore, the $0$-form generalization of this strong form would indeed place strong restrictions on axion moduli space diameters and close the aforementioned loophole.

However, even if we use this updated $N$-species strong form of the conjecture, there are other problems with invoking the strong WGC to rule out axion inflation.  To begin with, it does not rule out a closely related loophole achieved by taking $\mathcal{A}_1 < \mathcal{A}_2$ in (\ref{Veq}).  In this case, one could take $f_2$ arbitrarily large and $S_1 \lesssim S_2$ and still satisfy the strong WGC.  As long as $\mathcal{A}_1 e^{-S_1} \cos{a/f_1}$ is sufficiently smaller than $\mathcal{A}_2 e^{-S_2} \cos{a/f_2}$, the potential will be dominated by the latter term.  Secondly, it does not close the small action loophole discussed previously, in which the instanton actions are taken smaller than $1$.  This limit is difficult to arrange in a controlled string compactification, but as we have seen, it is not such a problem in simpler extranatural scenarios.  Most important, however, is the fact that the strong form of the WGC does not derive from arguments based on either effective field theory or black hole thermodynamics.  Though further developments could change the situation, we currently see no compelling reason to believe that the WGC holds in its stronger form.\footnote{See however \cite{Heidenreich:2015nta} for a discussion of the ``lattice WGC'' (LWGC), a candidate strong form which avoids some of these pitfalls and can be motivated by consistency considerations and string theory examples.}

\section{Conclusions}
\label{sec:conclude}

We have argued that the original magnetic form of the Weak Gravity Conjecture and the UV cutoff that it implies, appropriately generalized to multiple $U(1)$ gauge fields, exclude a variety of $N$-flation models including models of kinetic alignment. We summarize the claimed constraints, and the assumptions on which they rely, in Table \ref{tab:wgcresults}. The theories that are excluded in this way have in common the feature that the magnetic charges satisfying the magnetic WGC and the electric charges leading to the dominant instanton effects are simple (not parametrically large and aligned) in the same basis. We believe that these arguments are robust. They can possibly be evaded by considering a theory with a cancelation or tuning in the monopole masses (so that the monopoles are much lighter than the semiclassical self-energy estimate). The only other potential way out is if the compactification radius could somehow be consistently taken to be much smaller than the smallest distance $\Lambda^{-1}$ for which we trust the monopole solution. Because we expect the description of a local $U(1)$ gauge theory to break down at $\Lambda$, this would require going beyond the abelian effective field theory to a more complete ultraviolet description.

\afterpage{%
    \clearpage
    \thispagestyle{empty}
\begin{sidewaystable}
  \centering
  \begin{tabular}{l|l|l|l|l|l|l|l}
  Assumptions $\backslash$ Reference: & ref.~\cite{ArkaniHamed:2006dz} & ref.~\cite{delaFuente:2014aca} & ref.~\cite{Brown:2015lia} & sec.~\ref{sec:diagonalwarmup} & sec.~\ref{sec:magwgckineticalignment} & sec.~\ref{subsec:case1xwgc} & sec.~\ref{subsec:case2secc}  \\
  \hline
 Single axion & \cmark & \cmark & & & & & \\
 $K$ diagonal & \cmark & \cmark & WLOG & \cmark & & \cmark & \cmark  \\
 $S_{\rm inst} > 1$ & \cmark & & \cmark & & & & \\
 Instanton charges obey ElWGC  & \cmark & & \cmark & & & \cmark & \xmark \\
 MagWGC obeyed by simple charges  & & & & \cmark & \cmark & \cmark & \cmark \\
Instantons simple, $\sum a_i \cos(\theta_i)$  & & & & \cmark & \cmark & & \\
\hline
Constrained by: & ElWGC & MagWGC & ElWGC & MagWGC & MagWGC & ElXWGC +  & Single EFT + \\
 & & & & &  &  MagWGC & MagWGC \\
  \end{tabular}
  \caption{WGC constraints on (compact) axion inflation models with various assumptions. Each column is a scenario for which a constraint has been claimed; a \cmark~indicates that an assumption is made, and an \xmark~indicates that the {\em opposite} assumption is made. The entry ``WLOG'' for ``without loss of generality'' indicates that this assumption was made but, due to the lack of other related assumptions, it is a completely general basis choice. The single-axion assumption implies the $K$ diagonal assumption. The assumption that ``instanton charges obey the electric WGC'' misses the loophole where the states that satisfy the electric WGC make negligible contributions to the potential (for instance, they may have mass near the Planck scale, outside the effective theory). In theories with multiple axions the Convex Hull Condition is always assumed to be part of the definition of the appropriate WGC. Abbreviations: ElWGC = Electric Weak Gravity Conjecture, MagWGC = Magnetic Weak Gravity Conjecture, ElXWGC = Electric Extended Weak Gravity Conjecture.}
  \label{tab:wgcresults}
\end{sidewaystable}
\clearpage
}

A very interesting model in which the electric charge vectors for the instantons are highly aligned in the basis where magnetic charges are simple has been previously claimed to evade the Weak Gravity Conjecture \cite{delaFuente:2014aca}. We agree that it cannot be straightforwardly ruled out by the original WGC. However, this model implies surprising features that arise from the nontrivial dependence of the masses of charged particles on the values of the axion fields. We have proposed additional conjectures that would exclude such surprises. The two assumptions are that the mass spectrum at all extrema in the moduli space should satisfy the Weak Gravity Conjecture and that a light mode that is present in some region of moduli space should be part of the effective theory throughout the entire moduli space (rather than moving above the cutoff). We have no definitive proof of these statements, but they appear to be plausible, and we find it very compelling that they precisely parametrically exclude the one scenario that otherwise evades our arguments. Further study of these conjectures, as well as possible application of them to other models like axion monodromy inflation \cite{Silverstein:2008sg,McAllister:2008hb,Berg:2009tg,Kaloper:2011jz,Kaloper:2014zba} (with non-periodic potentials, unlike all cases considered in this paper), seem to us to be the most likely avenue for progress. We expect that the general argument sketched in section \ref{sec:monodromy} can exclude many such models, although they must be considered on a case-by-case basis to see if loopholes exist. Ultraviolet completions of other theories that apply large field ranges to the hierarchy problem \cite{Graham:2015cka} or to generating a light dilaton \cite{Contino2010,Bellazzini:2013fga,Coradeschi:2013gda} may be susceptible to similar constraints.

More generally, noncompact fields (like the scalar moduli which are supersymmetric counterparts of string axions) could give rise to large-field inflation, and the prospects for constraining them with the WGC are not clear. Still, it is thought that noncompact fields in string theory are highly constrained, and that effective field theory always breaks down in the presence of super-Planckian field ranges~\cite{Vafa:2005ui,Ooguri:2006in,Douglas:2005hq}. Super-Planckian field excursions in {\em space}, rather than in time, tend to collapse into black holes~\cite{Nicolis:2008wh}, which may point in the direction of general arguments against the consistency of effective field theories of super-Planckian fields coupled to gravity~\cite{Zohar}.

The Weak Gravity Conjecture has been established as a powerful tool to cull the space of theories of inflation. The possibility of future measurements of nonzero $r$ offers the hope that we can confront our understanding of general properties of quantum gravity against real empirical knowledge of our universe. We hope that the study of the WGC can offer tentative steps in the direction of a phenomenology of quantum gravity.

\section*{Acknowledgments}
We thank Thomas Bachlechner, Cliff Cheung, Thomas Dumitrescu, Peng Gao, Cody Long, Liam McAllister, Grant Remmen, Prashant Saraswat, Matt Schwartz, Gary Shiu, Eva Silverstein, Raman Sundrum, and Cumrun Vafa for discussions or correspondence. MR thanks Thomas Bachlechner, Liam McAllister, and Alexander Westphal for useful discussions after the first version of the paper, and in particular for suggesting the addition of Table \ref{tab:wgcresults}. BH is supported by the Fundamental Laws Initiative of the Harvard Center for the Fundamental Laws of Nature. The work of MR is supported in part by the NSF Grant PHY-1415548. TR is supported in part by the National Science Foundation under Grant No. DGE-1144152.  While preparing the revised version of this paper, MR was supported in part by the National Science Foundation under Grant No. PHYS-1066293 and the hospitality of the Aspen Center for Physics.

\appendix

\section{The SECC and gauge-invariant regulators}
\label{sec:lattice}

Our argument about the SECC relies, in part, on a discussion of short-wavelength modes of charged fields. In a gauge theory we expect that arbitrarily high-frequency gauge transformations should not be considered: gauge theories can be emergent long-distance descriptions of other physics, and the short distance degrees of freedom may be entirely different. However, because $\partial_\mu$ is not gauge invariant, while $D_\mu$ is, one may be concerned that the problem we are discussing is not actually a physical one. We believe that tracking physical (gauge invariant) masses of individual Kaluza--Klein modes adiabatically as the physical Wilson loop variable $\theta$ varies and seeing that they pass through the cutoff already demonstrates that the problem is physical. For completeness, in this appendix we provide a different perspective by looking at how two different gauge-invariant regulators treat the spectrum of modes across the moduli space.

Our first regulator is the lattice, with spacing $a = \Lambda^{-1}$. This regulator gives a good example in which 5d locality breaks down at short distances, but gauge invariance is preserved and we have a convenient setting to discuss only physical quantities. We will only discretize the fifth dimension, for simplicity. In other words, we will apply the idea of dimensional deconstruction \cite{ArkaniHamed:2001ca,Hill:2000mu,ArkaniHamed:2001nc,Hill:2002me}. We replace the continuous fifth dimension with a periodic lattice with $N$ sites. The SECC criterion of section \ref{sec:secc} tells us that, in order to work within the context of a single effective field theory valid everywhere in moduli space, we should take $N \gg q$. Let us explore what happens at smaller $N$ and how the spectrum alters as we take $N$ large. For a scalar field, we find a finite set of $N$ modes with masses
\be
m_n^2  =  \frac{N^2}{\pi^2 R^2} \sin^2\left(\frac{n \pi - q \theta/2}{N}\right), \label{eq:latticemasses}
\ee
with integer $n$ in the range $-N/2 < n \leq N/2$. In the large $N$ limit, for $n \ll N$, this reduces to the continuum Kaluza--Klein mode result $m_n^2 = \left(n - \frac{q\theta}{2 \pi}\right)^2/R^2$. Notice that the heaviest modes will have mass $\approx N/(\pi R)$, so to recover the continuum result for mode numbers up to $q$ we will need to take $N$ large compared to $\pi q$.

\begin{figure}[h]
\begin{center}
\includegraphics[width=0.8\textwidth]{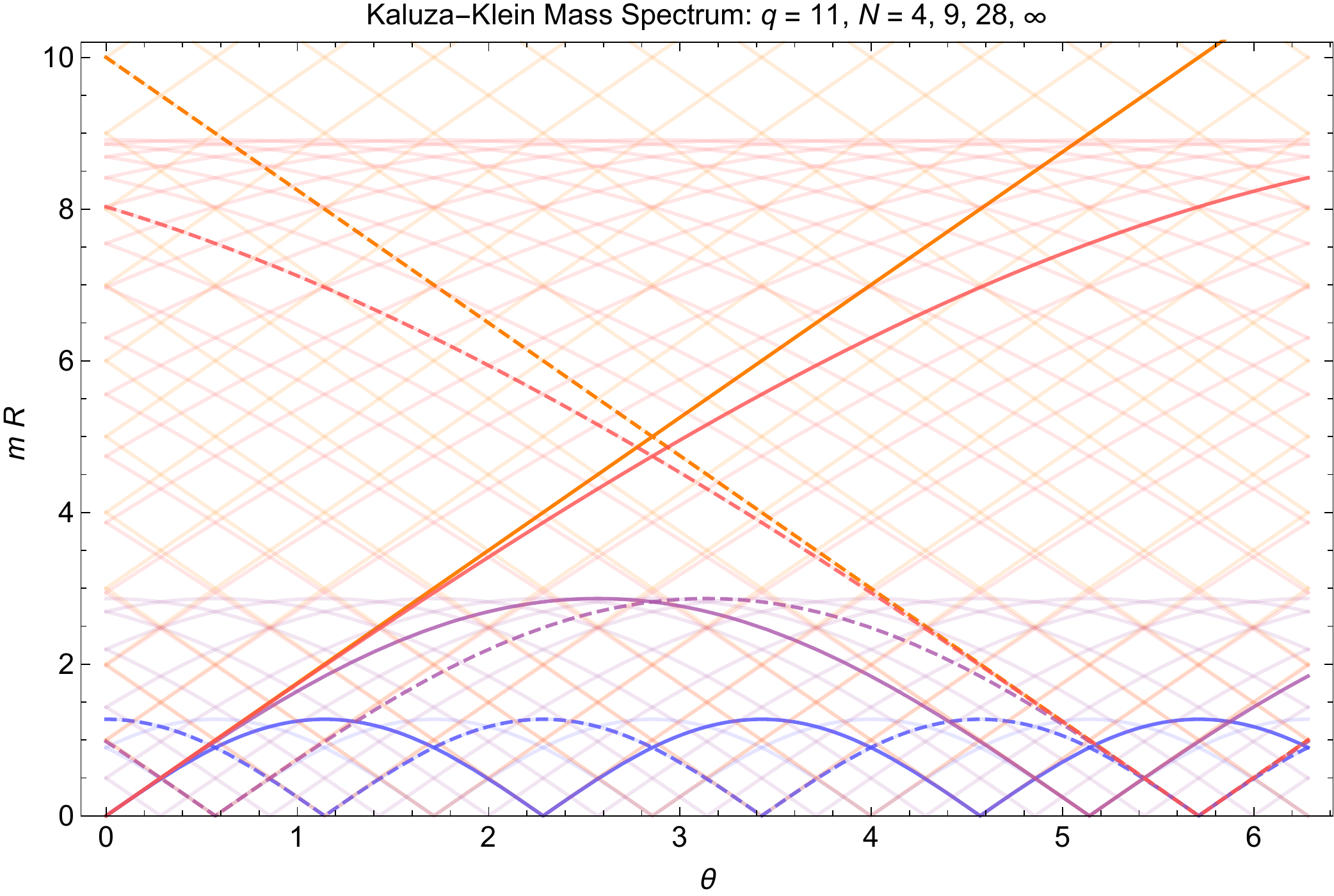}
\end{center}
\caption{Kaluza--Klein mode spectrum for a particle of charge $q = 11$ with an $N$-site lattice regulator for multiple choices of $N$. Blue curves are $N = 4$; purple, $N = 9$; red, $N = 28$; and orange, the continuum result $N \to \infty$. Observe that the spectrum of low-lying modes at $mR \simlt 1$ is the same for any number of lattice sites. However, the behavior of an individual mode tracked as $\theta$ is continuously varied is dramatically different at small $N$ and large $N$. The solid lines are modes with zero mass at the origin of moduli space, and the dashed lines are modes with zero mass at $\frac{\theta}{2\pi} = \frac{10}{11}$. Faint curves in the background show the other modes.}
\label{fig:kkspectrum}
\end{figure}

From the expression (\ref{eq:latticemasses}) we see that there is a zero mode whenever $\theta = 2\pi n/q$ (for integer $n$). This means that the lattice regulator brings down a light mode whenever the 5d continuum theory with a high cutoff would tell us to expect one, {\em independent of how small we take $N$}. One could take this as a hint that gauge invariance is subtle and any UV regulator will guarantee the consistency of the effective field theory. On the other hand, the origin of the light mode varies dramatically as we change $N$. In the continuum theory with a cutoff satisfying the SECC, the zero mode at $\theta = 2\pi n/q$ is different for each $n$. In the lattice theory at small $N$, there simply aren't enough modes for this to be true: a mode can be a zero mode, increase in mass as $\theta$ is varied, then turn around and decrease again to become a new zero mode at a different value of $\theta$. We illustrate this in figure \ref{fig:kkspectrum}. If the cutoff is sufficiently high, i.e. $N \gg q$, then a mode that becomes light at order-one values of $\theta$ was heavy at small $\theta$. When $N \ll q$, the modes that become light at any point in the moduli space were already light at the origin.

On the other hand, we could consider a different gauge invariant regulator. Suppose that we use a Pauli-Villars regulator, adding a mode with wrong-sign kinetic term and mass $\Lambda$ in the bulk. When $n-\frac{q\theta}{2\pi} \gg \Lambda R$, the contribution of a mode and its Pauli-Villars partner nearly cancel. In this case, in some sense all the modes are present for all $n$, but the regulator suppresses their contributions if they are not low-lying states. If we adiabatically track a mode that becomes light at large $\theta$, it will have been heavy at the origin, and vice versa. This is very different from the case of a lattice regulator, where individual modes periodically become light multiple times.

What we have found is that, for different gauge-invariant regulators with the same UV cutoff, the spectrum of light states will be similar but the behavior of individual modes as we move across the moduli space can be radically different. At the risk of belaboring the point, consider a physical state containing particles that are light at $\theta = 0$, and adiabatically vary $\theta$ to order-one values. The Pauli-Villars regulator would tell us that our particles keep becoming heavier. The lattice regulator would suggest that they would again become light. This is a sign of UV sensitivity and a breakdown of effective field theory. Only by taking a UV cutoff large enough to encompass at least $q$ modes at one time can we obtain a consistent answer that is insensitive to the regulator. Cutoffs that do not satisfy $q\theta \simlt \Lambda R$ require additional ultraviolet knowledge beyond the EFT. An interesting question to pursue further is just how mild this additional ultraviolet knowledge is: for instance, the lattice regulator discussed here is incompatible with knowing that the ultraviolet completion obeys Lorentz invariance in the extra dimension.

\bibliography{inflationref}

\providecommand{\href}[2]{#2}\begingroup\raggedright\begin{thebibliography}{10}

\bibitem{Lyth:1996im}
D.~H. Lyth, ``{What would we learn by detecting a gravitational wave signal in
  the cosmic microwave background anisotropy?},''
  \href{http://dx.doi.org/10.1103/PhysRevLett.78.1861}{{\em Phys.Rev.Lett.}
  {\bfseries 78} (1997) 1861--1863},
\href{http://arxiv.org/abs/hep-ph/9606387}{{\ttfamily arXiv:hep-ph/9606387
  [hep-ph]}}.

\bibitem{Easther:2006qu}
R.~Easther, W.~H. Kinney, and B.~A. Powell, ``{The Lyth bound and the end of
  inflation},'' \href{http://dx.doi.org/10.1088/1475-7516/2006/08/004}{{\em
  JCAP} {\bfseries 0608} (2006) 004},
\href{http://arxiv.org/abs/astro-ph/0601276}{{\ttfamily arXiv:astro-ph/0601276
  [astro-ph]}}.

\bibitem{Baumann:2011ws}
D.~Baumann and D.~Green, ``{A Field Range Bound for General Single-Field
  Inflation},'' \href{http://dx.doi.org/10.1088/1475-7516/2012/05/017}{{\em
  JCAP} {\bfseries 1205} (2012) 017},
\href{http://arxiv.org/abs/1111.3040}{{\ttfamily arXiv:1111.3040 [hep-th]}}.

\bibitem{Antusch:2014cpa}
S.~Antusch and D.~Nolde, ``{BICEP2 implications for single-field slow-roll
  inflation revisited},''
  \href{http://dx.doi.org/10.1088/1475-7516/2014/05/035}{{\em JCAP} {\bfseries
  1405} (2014) 035},
\href{http://arxiv.org/abs/1404.1821}{{\ttfamily arXiv:1404.1821 [hep-ph]}}.

\bibitem{Bramante:2014rva}
J.~Bramante, S.~Downes, L.~Lehman, and A.~Martin, ``{Last stand of single small
  field inflation},'' \href{http://dx.doi.org/10.1103/PhysRevD.90.023530}{{\em
  Phys.Rev.} {\bfseries D90} no.~2, (2014) 023530},
\href{http://arxiv.org/abs/1405.7563}{{\ttfamily arXiv:1405.7563
  [astro-ph.CO]}}.

\bibitem{Kamionkowski:1992mf}
M.~Kamionkowski and J.~March-Russell, ``{Planck scale physics and the
  Peccei-Quinn mechanism},''
  \href{http://dx.doi.org/10.1016/0370-2693(92)90492-M}{{\em Phys.Lett.}
  {\bfseries B282} (1992) 137--141},
\href{http://arxiv.org/abs/hep-th/9202003}{{\ttfamily arXiv:hep-th/9202003
  [hep-th]}}.

\bibitem{Holman:1992us}
R.~Holman, S.~D. Hsu, T.~W. Kephart, E.~W. Kolb, R.~Watkins, and L.~M. Widrow,
  ``{Solutions to the strong CP problem in a world with gravity},''
  \href{http://dx.doi.org/10.1016/0370-2693(92)90491-L}{{\em Phys.Lett.}
  {\bfseries B282} (1992) 132--136},
\href{http://arxiv.org/abs/hep-ph/9203206}{{\ttfamily arXiv:hep-ph/9203206
  [hep-ph]}}.

\bibitem{Kallosh:1995hi}
R.~Kallosh, A.~D. Linde, D.~A. Linde, and L.~Susskind, ``{Gravity and global
  symmetries},'' \href{http://dx.doi.org/10.1103/PhysRevD.52.912}{{\em
  Phys.Rev.} {\bfseries D52} (1995) 912--935},
\href{http://arxiv.org/abs/hep-th/9502069}{{\ttfamily arXiv:hep-th/9502069
  [hep-th]}}.

\bibitem{Conlon:2006tq}
J.~P. Conlon, ``{The QCD axion and moduli stabilisation},''
  \href{http://dx.doi.org/10.1088/1126-6708/2006/05/078}{{\em JHEP} {\bfseries
  0605} (2006) 078},
\href{http://arxiv.org/abs/hep-th/0602233}{{\ttfamily arXiv:hep-th/0602233
  [hep-th]}}.

\bibitem{Svrcek:2006yi}
P.~Svrcek and E.~Witten, ``{Axions In String Theory},''
  \href{http://dx.doi.org/10.1088/1126-6708/2006/06/051}{{\em JHEP} {\bfseries
  0606} (2006) 051},
\href{http://arxiv.org/abs/hep-th/0605206}{{\ttfamily arXiv:hep-th/0605206
  [hep-th]}}.

\bibitem{Freese:1990rb}
K.~Freese, J.~A. Frieman, and A.~V. Olinto, ``{Natural inflation with pseudo -
  Nambu-Goldstone bosons},''
\href{http://dx.doi.org/10.1103/PhysRevLett.65.3233}{{\em Phys.Rev.Lett.}
  {\bfseries 65} (1990) 3233--3236}.

\bibitem{Banks:1995dp}
T.~Banks, M.~Berkooz, S.~Shenker, G.~W. Moore, and P.~Steinhardt, ``{Modular
  cosmology},'' \href{http://dx.doi.org/10.1103/PhysRevD.52.3548}{{\em
  Phys.Rev.} {\bfseries D52} (1995) 3548--3562},
\href{http://arxiv.org/abs/hep-th/9503114}{{\ttfamily arXiv:hep-th/9503114
  [hep-th]}}.

\bibitem{ArkaniHamed:2003wu}
N.~Arkani-Hamed, H.-C. Cheng, P.~Creminelli, and L.~Randall, ``{Extra natural
  inflation},'' \href{http://dx.doi.org/10.1103/PhysRevLett.90.221302}{{\em
  Phys.Rev.Lett.} {\bfseries 90} (2003) 221302},
\href{http://arxiv.org/abs/hep-th/0301218}{{\ttfamily arXiv:hep-th/0301218
  [hep-th]}}.

\bibitem{Kim:2004rp}
J.~E. Kim, H.~P. Nilles, and M.~Peloso, ``{Completing natural inflation},''
  \href{http://dx.doi.org/10.1088/1475-7516/2005/01/005}{{\em JCAP} {\bfseries
  0501} (2005) 005},
\href{http://arxiv.org/abs/hep-ph/0409138}{{\ttfamily arXiv:hep-ph/0409138
  [hep-ph]}}.

\bibitem{Dimopoulos:2005ac}
S.~Dimopoulos, S.~Kachru, J.~McGreevy, and J.~G. Wacker, ``{N-flation},''
  \href{http://dx.doi.org/10.1088/1475-7516/2008/08/003}{{\em JCAP} {\bfseries
  0808} (2008) 003},
\href{http://arxiv.org/abs/hep-th/0507205}{{\ttfamily arXiv:hep-th/0507205
  [hep-th]}}.

\bibitem{Silverstein:2008sg}
E.~Silverstein and A.~Westphal, ``{Monodromy in the CMB: Gravity Waves and
  String Inflation},'' \href{http://dx.doi.org/10.1103/PhysRevD.78.106003}{{\em
  Phys.Rev.} {\bfseries D78} (2008) 106003},
\href{http://arxiv.org/abs/0803.3085}{{\ttfamily arXiv:0803.3085 [hep-th]}}.

\bibitem{McAllister:2008hb}
L.~McAllister, E.~Silverstein, and A.~Westphal, ``{Gravity Waves and Linear
  Inflation from Axion Monodromy},''
  \href{http://dx.doi.org/10.1103/PhysRevD.82.046003}{{\em Phys.Rev.}
  {\bfseries D82} (2010) 046003},
\href{http://arxiv.org/abs/0808.0706}{{\ttfamily arXiv:0808.0706 [hep-th]}}.

\bibitem{Berg:2009tg}
M.~Berg, E.~Pajer, and S.~Sjors, ``{Dante's Inferno},''
  \href{http://dx.doi.org/10.1103/PhysRevD.81.103535}{{\em Phys.Rev.}
  {\bfseries D81} (2010) 103535},
\href{http://arxiv.org/abs/0912.1341}{{\ttfamily arXiv:0912.1341 [hep-th]}}.

\bibitem{Germani:2010hd}
C.~Germani and A.~Kehagias, ``{UV-Protected Inflation},''
  \href{http://dx.doi.org/10.1103/PhysRevLett.106.161302}{{\em Phys. Rev.
  Lett.} {\bfseries 106} (2011) 161302},
\href{http://arxiv.org/abs/1012.0853}{{\ttfamily arXiv:1012.0853 [hep-ph]}}.

\bibitem{Kaloper:2011jz}
N.~Kaloper, A.~Lawrence, and L.~Sorbo, ``{An Ignoble Approach to Large Field
  Inflation},'' \href{http://dx.doi.org/10.1088/1475-7516/2011/03/023}{{\em
  JCAP} {\bfseries 1103} (2011) 023},
\href{http://arxiv.org/abs/1101.0026}{{\ttfamily arXiv:1101.0026 [hep-th]}}.

\bibitem{Kaloper:2014zba}
N.~Kaloper and A.~Lawrence, ``{Natural chaotic inflation and ultraviolet
  sensitivity},'' \href{http://dx.doi.org/10.1103/PhysRevD.90.023506}{{\em
  Phys.Rev.} {\bfseries D90} no.~2, (2014) 023506},
\href{http://arxiv.org/abs/1404.2912}{{\ttfamily arXiv:1404.2912 [hep-th]}}.

\bibitem{Burgess:2014tja}
C.~P. Burgess, M.~Cicoli, F.~Quevedo, and M.~Williams, ``{Inflating with Large
  Effective Fields},''
  \href{http://dx.doi.org/10.1088/1475-7516/2014/11/045}{{\em JCAP} {\bfseries
  1411} (2014) 045},
\href{http://arxiv.org/abs/1404.6236}{{\ttfamily arXiv:1404.6236 [hep-th]}}.

\bibitem{Bachlechner:2014hsa}
T.~C. Bachlechner, M.~Dias, J.~Frazer, and L.~McAllister, ``{Chaotic inflation
  with kinetic alignment of axion fields},''
  \href{http://dx.doi.org/10.1103/PhysRevD.91.023520}{{\em Phys.Rev.}
  {\bfseries D91} no.~2, (2015) 023520},
\href{http://arxiv.org/abs/1404.7496}{{\ttfamily arXiv:1404.7496 [hep-th]}}.

\bibitem{Csaki:2014bua}
C.~Csaki, N.~Kaloper, J.~Serra, and J.~Terning, ``{Inflation from Broken Scale
  Invariance},'' \href{http://dx.doi.org/10.1103/PhysRevLett.113.161302}{{\em
  Phys.Rev.Lett.} {\bfseries 113} (2014) 161302},
\href{http://arxiv.org/abs/1406.5192}{{\ttfamily arXiv:1406.5192 [hep-th]}}.

\bibitem{Furuuchi:2014cwa}
K.~Furuuchi and Y.~Koyama, ``{Large field inflation models from
  higher-dimensional gauge theories},''
  \href{http://dx.doi.org/10.1088/1475-7516/2015/02/031}{{\em JCAP} {\bfseries
  1502} no.~02, (2015) 031},
\href{http://arxiv.org/abs/1407.1951}{{\ttfamily arXiv:1407.1951 [hep-th]}}.

\bibitem{Harigaya:2014rga}
K.~Harigaya and M.~Ibe, ``{Phase Locked Inflation. Effectively Trans-Planckian
  Natural Inflation},'' \href{http://dx.doi.org/10.1007/JHEP11(2014)147}{{\em
  JHEP} {\bfseries 1411} (2014) 147},
\href{http://arxiv.org/abs/1407.4893}{{\ttfamily arXiv:1407.4893 [hep-ph]}}.

\bibitem{Higaki:2014mwa}
T.~Higaki and F.~Takahashi, ``{Axion Landscape and Natural Inflation},''
  \href{http://dx.doi.org/10.1016/j.physletb.2015.03.052}{{\em Phys.Lett.}
  {\bfseries B744} (2015) 153--159},
\href{http://arxiv.org/abs/1409.8409}{{\ttfamily arXiv:1409.8409 [hep-ph]}}.

\bibitem{Shiu:2015xda}
G.~Shiu, W.~Staessens, and F.~Ye, ``{Large Field Inflation from Axion
  Mixing},'' \href{http://dx.doi.org/10.1007/JHEP06(2015)026}{{\em JHEP}
  {\bfseries 06} (2015) 026},
\href{http://arxiv.org/abs/1503.02965}{{\ttfamily arXiv:1503.02965 [hep-th]}}.

\bibitem{Liddle:1998jc}
A.~R. Liddle, A.~Mazumdar, and F.~E. Schunck, ``{Assisted inflation},''
  \href{http://dx.doi.org/10.1103/PhysRevD.58.061301}{{\em Phys. Rev.}
  {\bfseries D58} (1998) 061301},
\href{http://arxiv.org/abs/astro-ph/9804177}{{\ttfamily arXiv:astro-ph/9804177
  [astro-ph]}}.

\bibitem{Copeland:1999cs}
E.~J. Copeland, A.~Mazumdar, and N.~J. Nunes, ``{Generalized assisted
  inflation},'' \href{http://dx.doi.org/10.1103/PhysRevD.60.083506}{{\em Phys.
  Rev.} {\bfseries D60} (1999) 083506},
\href{http://arxiv.org/abs/astro-ph/9904309}{{\ttfamily arXiv:astro-ph/9904309
  [astro-ph]}}.

\bibitem{Mazumdar:2001mm}
A.~Mazumdar, S.~Panda, and A.~Perez-Lorenzana, ``{Assisted inflation via
  tachyon condensation},''
  \href{http://dx.doi.org/10.1016/S0550-3213(01)00410-2}{{\em Nucl. Phys.}
  {\bfseries B614} (2001) 101--116},
\href{http://arxiv.org/abs/hep-ph/0107058}{{\ttfamily arXiv:hep-ph/0107058
  [hep-ph]}}.

\bibitem{Jokinen:2004bp}
A.~Jokinen and A.~Mazumdar, ``{Inflation in large N limit of supersymmetric
  gauge theories},''
  \href{http://dx.doi.org/10.1016/j.physletb.2004.07.010}{{\em Phys. Lett.}
  {\bfseries B597} (2004) 222},
\href{http://arxiv.org/abs/hep-th/0406074}{{\ttfamily arXiv:hep-th/0406074
  [hep-th]}}.

\bibitem{Ashoorioon:2009wa}
A.~Ashoorioon, H.~Firouzjahi, and M.~M. Sheikh-Jabbari, ``{M-flation: Inflation
  From Matrix Valued Scalar Fields},''
  \href{http://dx.doi.org/10.1088/1475-7516/2009/06/018}{{\em JCAP} {\bfseries
  0906} (2009) 018},
\href{http://arxiv.org/abs/0903.1481}{{\ttfamily arXiv:0903.1481 [hep-th]}}.

\bibitem{Ashoorioon:2011ki}
A.~Ashoorioon and M.~M. Sheikh-Jabbari, ``{Gauged M-flation, its UV sensitivity
  and Spectator Species},''
  \href{http://dx.doi.org/10.1088/1475-7516/2011/06/014}{{\em JCAP} {\bfseries
  1106} (2011) 014},
\href{http://arxiv.org/abs/1101.0048}{{\ttfamily arXiv:1101.0048 [hep-th]}}.

\bibitem{Ashoorioon:2014jja}
A.~Ashoorioon and M.~M. Sheikh-Jabbari, ``{Gauged M-flation After BICEP2},''
  \href{http://dx.doi.org/10.1016/j.physletb.2014.11.018}{{\em Phys. Lett.}
  {\bfseries B739} (2014) 391--399},
\href{http://arxiv.org/abs/1405.1685}{{\ttfamily arXiv:1405.1685 [hep-th]}}.

\bibitem{Banks:2003sx}
T.~Banks, M.~Dine, P.~J. Fox, and E.~Gorbatov, ``{On the possibility of large
  axion decay constants},''
  \href{http://dx.doi.org/10.1088/1475-7516/2003/06/001}{{\em JCAP} {\bfseries
  0306} (2003) 001},
\href{http://arxiv.org/abs/hep-th/0303252}{{\ttfamily arXiv:hep-th/0303252
  [hep-th]}}.

\bibitem{Rudelius:2014wla}
T.~Rudelius, ``{On the Possibility of Large Axion Moduli Spaces},''
  \href{http://dx.doi.org/10.1088/1475-7516/2015/04/049}{{\em JCAP} {\bfseries
  1504} no.~04, (2015) 049},
\href{http://arxiv.org/abs/1409.5793}{{\ttfamily arXiv:1409.5793 [hep-th]}}.

\bibitem{Vafa:2005ui}
C.~Vafa, ``{The String landscape and the swampland},''
\href{http://arxiv.org/abs/hep-th/0509212}{{\ttfamily arXiv:hep-th/0509212
  [hep-th]}}.

\bibitem{ArkaniHamed:2006dz}
N.~Arkani-Hamed, L.~Motl, A.~Nicolis, and C.~Vafa, ``{The String landscape,
  black holes and gravity as the weakest force},''
  \href{http://dx.doi.org/10.1088/1126-6708/2007/06/060}{{\em JHEP} {\bfseries
  0706} (2007) 060},
\href{http://arxiv.org/abs/hep-th/0601001}{{\ttfamily arXiv:hep-th/0601001
  [hep-th]}}.

\bibitem{Ooguri:2006in}
H.~Ooguri and C.~Vafa, ``{On the Geometry of the String Landscape and the
  Swampland},'' \href{http://dx.doi.org/10.1016/j.nuclphysb.2006.10.033}{{\em
  Nucl.Phys.} {\bfseries B766} (2007) 21--33},
\href{http://arxiv.org/abs/hep-th/0605264}{{\ttfamily arXiv:hep-th/0605264
  [hep-th]}}.

\bibitem{Douglas:2005hq}
M.~R. Douglas and Z.~Lu, ``{Finiteness of volume of moduli spaces},''
\href{http://arxiv.org/abs/hep-th/0509224}{{\ttfamily arXiv:hep-th/0509224
  [hep-th]}}.

\bibitem{Kats:2006xp}
Y.~Kats, L.~Motl, and M.~Padi, ``{Higher-order corrections to mass-charge
  relation of extremal black holes},''
  \href{http://dx.doi.org/10.1088/1126-6708/2007/12/068}{{\em JHEP} {\bfseries
  0712} (2007) 068},
\href{http://arxiv.org/abs/hep-th/0606100}{{\ttfamily arXiv:hep-th/0606100
  [hep-th]}}.

\bibitem{Banks:2006mm}
T.~Banks, M.~Johnson, and A.~Shomer, ``{A Note on Gauge Theories Coupled to
  Gravity},'' \href{http://dx.doi.org/10.1088/1126-6708/2006/09/049}{{\em JHEP}
  {\bfseries 0609} (2006) 049},
\href{http://arxiv.org/abs/hep-th/0606277}{{\ttfamily arXiv:hep-th/0606277
  [hep-th]}}.

\bibitem{Cheung:2014vva}
C.~Cheung and G.~N. Remmen, ``{Naturalness and the Weak Gravity Conjecture},''
  \href{http://dx.doi.org/10.1103/PhysRevLett.113.051601}{{\em Phys.Rev.Lett.}
  {\bfseries 113} (2014) 051601},
\href{http://arxiv.org/abs/1402.2287}{{\ttfamily arXiv:1402.2287 [hep-ph]}}.

\bibitem{Cheung:2014ega}
C.~Cheung and G.~N. Remmen, ``{Infrared Consistency and the Weak Gravity
  Conjecture},'' \href{http://dx.doi.org/10.1007/JHEP12(2014)087}{{\em JHEP}
  {\bfseries 1412} (2014) 087},
\href{http://arxiv.org/abs/1407.7865}{{\ttfamily arXiv:1407.7865 [hep-th]}}.

\bibitem{Susskind:1995da}
L.~Susskind, ``{Trouble for remnants},''
\href{http://arxiv.org/abs/hep-th/9501106}{{\ttfamily arXiv:hep-th/9501106
  [hep-th]}}.

\bibitem{Bachlechner:2014gfa}
T.~C. Bachlechner, C.~Long, and L.~McAllister, ``{Planckian Axions in String
  Theory},''
\href{http://arxiv.org/abs/1412.1093}{{\ttfamily arXiv:1412.1093 [hep-th]}}.

\bibitem{delaFuente:2014aca}
A.~de~la Fuente, P.~Saraswat, and R.~Sundrum, ``{Natural Inflation and Quantum
  Gravity},'' \href{http://dx.doi.org/10.1103/PhysRevLett.114.151303}{{\em
  Phys.Rev.Lett.} {\bfseries 114} no.~15, (2015) 151303},
\href{http://arxiv.org/abs/1412.3457}{{\ttfamily arXiv:1412.3457 [hep-th]}}.

\bibitem{Rudelius:2015xta}
T.~Rudelius, ``{Constraints on Axion Inflation from the Weak Gravity
  Conjecture},'' \href{http://dx.doi.org/10.1088/1475-7516/2015/9/020}{{\em
  JCAP} {\bfseries 09} (2015) 020},
\href{http://arxiv.org/abs/1503.00795}{{\ttfamily arXiv:1503.00795 [hep-th]}}.

\bibitem{Montero:2015ofa}
M.~Montero, A.~M. Uranga, and I.~Valenzuela, ``{Transplanckian axions!?},''
  \href{http://dx.doi.org/10.1007/JHEP08(2015)032}{{\em JHEP} {\bfseries 08}
  (2015) 032},
\href{http://arxiv.org/abs/1503.03886}{{\ttfamily arXiv:1503.03886 [hep-th]}}.

\bibitem{Brown:2015iha}
J.~Brown, W.~Cottrell, G.~Shiu, and P.~Soler, ``{Fencing in the Swampland:
  Quantum Gravity Constraints on Large Field Inflation},''
\href{http://arxiv.org/abs/1503.04783}{{\ttfamily arXiv:1503.04783 [hep-th]}}.

\bibitem{Bachlechner:2015qja}
T.~C. Bachlechner, C.~Long, and L.~McAllister, ``{Planckian Axions and the Weak
  Gravity Conjecture},''
\href{http://arxiv.org/abs/1503.07853}{{\ttfamily arXiv:1503.07853 [hep-th]}}.

\bibitem{Hebecker:2015rya}
A.~Hebecker, P.~Mangat, F.~Rompineve, and L.~T. Witkowski, ``{Winding out of
  the Swamp: Evading the Weak Gravity Conjecture with F-term Winding
  Inflation?},'' \href{http://dx.doi.org/10.1016/j.physletb.2015.07.026}{{\em
  Phys. Lett.} {\bfseries B748} (2015) 455--462},
\href{http://arxiv.org/abs/1503.07912}{{\ttfamily arXiv:1503.07912 [hep-th]}}.

\bibitem{Brown:2015lia}
J.~Brown, W.~Cottrell, G.~Shiu, and P.~Soler, ``{On Axionic Field Ranges,
  Loopholes and the Weak Gravity Conjecture},''
\href{http://arxiv.org/abs/1504.00659}{{\ttfamily arXiv:1504.00659 [hep-th]}}.

\bibitem{Junghans:2015hba}
D.~Junghans, ``{Large-Field Inflation with Multiple Axions and the Weak Gravity
  Conjecture},''
\href{http://arxiv.org/abs/1504.03566}{{\ttfamily arXiv:1504.03566 [hep-th]}}.

\bibitem{Heidenreich:2015nta}
B.~Heidenreich, M.~Reece, and T.~Rudelius, ``{Sharpening the Weak Gravity
  Conjecture with Dimensional Reduction},''
\href{http://arxiv.org/abs/1509.06374}{{\ttfamily arXiv:1509.06374 [hep-th]}}.

\bibitem{Hiscock:1990ex}
W.~A. Hiscock and L.~D. Weems, ``{Evolution of Charged Evaporating Black
  Holes},''
\href{http://dx.doi.org/10.1103/PhysRevD.41.1142}{{\em Phys.Rev.} {\bfseries
  D41} (1990) 1142}.

\bibitem{Gibbons:1975kk}
G.~Gibbons, ``{Vacuum Polarization and the Spontaneous Loss of Charge by Black
  Holes},''
\href{http://dx.doi.org/10.1007/BF01609829}{{\em Commun.Math.Phys.} {\bfseries
  44} (1975) 245--264}.

\bibitem{Schumacher:1985zz}
B.~Schumacher, ``{Is Bekenstein's Conjecture True for Charged Black Holes?},''
\href{http://dx.doi.org/10.1103/PhysRevLett.54.2643}{{\em Phys.Rev.Lett.}
  {\bfseries 54} (1985) 2643--2645}.

\bibitem{Khriplovich:1999gm}
I.~Khriplovich, ``{Particle creation by charged black holes},''
\href{http://dx.doi.org/10.1016/S0370-1573(99)00078-2}{{\em Phys.Rept.}
  {\bfseries 320} (1999) 37--49}.

\bibitem{Denef:2009tp}
F.~Denef and S.~A. Hartnoll, ``{Landscape of superconducting membranes},''
  \href{http://dx.doi.org/10.1103/PhysRevD.79.126008}{{\em Phys.Rev.}
  {\bfseries D79} (2009) 126008},
\href{http://arxiv.org/abs/0901.1160}{{\ttfamily arXiv:0901.1160 [hep-th]}}.

\bibitem{Chen:2012zn}
C.-M. Chen, S.~P. Kim, I.-C. Lin, J.-R. Sun, and M.-F. Wu, ``{Spontaneous Pair
  Production in Reissner-Nordstrom Black Holes},''
  \href{http://dx.doi.org/10.1103/PhysRevD.85.124041}{{\em Phys.Rev.}
  {\bfseries D85} (2012) 124041},
\href{http://arxiv.org/abs/1202.3224}{{\ttfamily arXiv:1202.3224 [hep-th]}}.

\bibitem{Chen:2014yfa}
C.-M. Chen, J.-R. Sun, F.-Y. Tang, and P.-Y. Tsai, ``{Spinor particle creation
  in near extremal Reissner-Nordstr{\"o}m black holes},''
  \href{http://dx.doi.org/10.1088/0264-9381/32/19/195003}{{\em Class. Quant.
  Grav.} {\bfseries 32} no.~19, (2015) 195003},
\href{http://arxiv.org/abs/1412.6876}{{\ttfamily arXiv:1412.6876 [hep-th]}}.

\bibitem{Cornell2015}
T.~C. Bachlechner, C.~Long, L.~McAllister, and J.~Stout, 2015.
\newblock {To appear}.

\bibitem{Prashant}
A.~de~la Fuente, P.~Saraswat, and R.~Sundrum, 2014.
\newblock {Private communication}.

\bibitem{Kaloper:1999tt}
N.~Kaloper and A.~D. Linde, ``{Cosmology versus holography},''
  \href{http://dx.doi.org/10.1103/PhysRevD.60.103509}{{\em Phys.Rev.}
  {\bfseries D60} (1999) 103509},
\href{http://arxiv.org/abs/hep-th/9904120}{{\ttfamily arXiv:hep-th/9904120
  [hep-th]}}.

\bibitem{Conlon:2012tz}
J.~P. Conlon, ``{Quantum Gravity Constraints on Inflation},''
  \href{http://dx.doi.org/10.1088/1475-7516/2012/09/019}{{\em JCAP} {\bfseries
  1209} (2012) 019},
\href{http://arxiv.org/abs/1203.5476}{{\ttfamily arXiv:1203.5476 [hep-th]}}.

\bibitem{Boubekeur:2013kga}
L.~Boubekeur, ``{On the Scale of New Physics in Inflation},''
\href{http://arxiv.org/abs/1312.4768}{{\ttfamily arXiv:1312.4768
  [astro-ph.CO]}}.

\bibitem{Hosotani:1983xw}
Y.~Hosotani, ``{Dynamical Mass Generation by Compact Extra Dimensions},''
\href{http://dx.doi.org/10.1016/0370-2693(83)90170-3}{{\em Phys.Lett.}
  {\bfseries B126} (1983) 309}.

\bibitem{Cheng:2002iz}
H.-C. Cheng, K.~T. Matchev, and M.~Schmaltz, ``{Radiative corrections to
  Kaluza-Klein masses},''
  \href{http://dx.doi.org/10.1103/PhysRevD.66.036005}{{\em Phys.Rev.}
  {\bfseries D66} (2002) 036005},
\href{http://arxiv.org/abs/hep-ph/0204342}{{\ttfamily arXiv:hep-ph/0204342
  [hep-ph]}}.

\bibitem{ArkaniHamed:2007gg}
N.~Arkani-Hamed, S.~Dubovsky, A.~Nicolis, and G.~Villadoro, ``{Quantum Horizons
  of the Standard Model Landscape},''
  \href{http://dx.doi.org/10.1088/1126-6708/2007/06/078}{{\em JHEP} {\bfseries
  0706} (2007) 078},
\href{http://arxiv.org/abs/hep-th/0703067}{{\ttfamily arXiv:hep-th/0703067
  [HEP-TH]}}.

\bibitem{Seiberg:1994rs}
N.~Seiberg and E.~Witten, ``{Electric - magnetic duality, monopole
  condensation, and confinement in N=2 supersymmetric Yang-Mills theory},''
  \href{http://dx.doi.org/10.1016/0550-3213(94)90124-4}{{\em Nucl.Phys.}
  {\bfseries B426} (1994) 19--52},
\href{http://arxiv.org/abs/hep-th/9407087}{{\ttfamily arXiv:hep-th/9407087
  [hep-th]}}.

\bibitem{Bekenstein:1980jp}
J.~D. Bekenstein, ``{A Universal Upper Bound on the Entropy to Energy Ratio for
  Bounded Systems},''
\href{http://dx.doi.org/10.1103/PhysRevD.23.287}{{\em Phys.Rev.} {\bfseries
  D23} (1981) 287}.

\bibitem{Bousso:1999xy}
R.~Bousso, ``{A Covariant entropy conjecture},''
  \href{http://dx.doi.org/10.1088/1126-6708/1999/07/004}{{\em JHEP} {\bfseries
  9907} (1999) 004},
\href{http://arxiv.org/abs/hep-th/9905177}{{\ttfamily arXiv:hep-th/9905177
  [hep-th]}}.

\bibitem{Srednicki:1993im}
M.~Srednicki, ``{Entropy and area},''
  \href{http://dx.doi.org/10.1103/PhysRevLett.71.666}{{\em Phys.Rev.Lett.}
  {\bfseries 71} (1993) 666--669},
\href{http://arxiv.org/abs/hep-th/9303048}{{\ttfamily arXiv:hep-th/9303048
  [hep-th]}}.

\bibitem{Casini:2009sr}
H.~Casini and M.~Huerta, ``{Entanglement entropy in free quantum field
  theory},'' \href{http://dx.doi.org/10.1088/1751-8113/42/50/504007}{{\em
  J.Phys.} {\bfseries A42} (2009) 504007},
\href{http://arxiv.org/abs/0905.2562}{{\ttfamily arXiv:0905.2562 [hep-th]}}.

\bibitem{Casini:2008cr}
H.~Casini, ``{Relative entropy and the Bekenstein bound},''
  \href{http://dx.doi.org/10.1088/0264-9381/25/20/205021}{{\em
  Class.Quant.Grav.} {\bfseries 25} (2008) 205021},
\href{http://arxiv.org/abs/0804.2182}{{\ttfamily arXiv:0804.2182 [hep-th]}}.

\bibitem{Bousso:2014sda}
R.~Bousso, H.~Casini, Z.~Fisher, and J.~Maldacena, ``{Proof of a Quantum Bousso
  Bound},'' \href{http://dx.doi.org/10.1103/PhysRevD.90.044002}{{\em Phys.Rev.}
  {\bfseries D90} no.~4, (2014) 044002},
\href{http://arxiv.org/abs/1404.5635}{{\ttfamily arXiv:1404.5635 [hep-th]}}.

\bibitem{Bousso:2014uxa}
R.~Bousso, H.~Casini, Z.~Fisher, and J.~Maldacena, ``{Entropy on a null surface
  for interacting quantum field theories and the Bousso bound},''
  \href{http://dx.doi.org/10.1103/PhysRevD.91.084030}{{\em Phys.Rev.}
  {\bfseries D91} no.~8, (2015) 084030},
\href{http://arxiv.org/abs/1406.4545}{{\ttfamily arXiv:1406.4545 [hep-th]}}.

\bibitem{Gibbons:1977mu}
G.~Gibbons and S.~Hawking, ``{Cosmological Event Horizons, Thermodynamics, and
  Particle Creation},''
\href{http://dx.doi.org/10.1103/PhysRevD.15.2738}{{\em Phys.Rev.} {\bfseries
  D15} (1977) 2738--2751}.

\bibitem{Frolov:2002va}
A.~V. Frolov and L.~Kofman, ``{Inflation and de Sitter thermodynamics},''
  \href{http://dx.doi.org/10.1088/1475-7516/2003/05/009}{{\em JCAP} {\bfseries
  0305} (2003) 009},
\href{http://arxiv.org/abs/hep-th/0212327}{{\ttfamily arXiv:hep-th/0212327
  [hep-th]}}.

\bibitem{Graham:2015cka}
P.~W. Graham, D.~E. Kaplan, and S.~Rajendran, ``{Cosmological Relaxation of the
  Electroweak Scale},''
\href{http://arxiv.org/abs/1504.07551}{{\ttfamily arXiv:1504.07551 [hep-ph]}}.

\bibitem{Contino2010}
R.~Contino, A.~Pomarol, and R.~Rattazzi, 2010.
\newblock {Unpublished work discussed in talks at Xmas10 and Planck 2010}.

\bibitem{Bellazzini:2013fga}
B.~Bellazzini, C.~Csaki, J.~Hubisz, J.~Serra, and J.~Terning, ``{A Naturally
  Light Dilaton and a Small Cosmological Constant},''
  \href{http://dx.doi.org/10.1140/epjc/s10052-014-2790-x}{{\em Eur.Phys.J.}
  {\bfseries C74} (2014) 2790},
\href{http://arxiv.org/abs/1305.3919}{{\ttfamily arXiv:1305.3919 [hep-th]}}.

\bibitem{Coradeschi:2013gda}
F.~Coradeschi, P.~Lodone, D.~Pappadopulo, R.~Rattazzi, and L.~Vitale, ``{A
  naturally light dilaton},''
  \href{http://dx.doi.org/10.1007/JHEP11(2013)057}{{\em JHEP} {\bfseries 1311}
  (2013) 057},
\href{http://arxiv.org/abs/1306.4601}{{\ttfamily arXiv:1306.4601 [hep-th]}}.

\bibitem{Nicolis:2008wh}
A.~Nicolis, ``{On Super-Planckian Fields at Sub-Planckian Energies},''
  \href{http://dx.doi.org/10.1088/1126-6708/2008/07/023}{{\em JHEP} {\bfseries
  0807} (2008) 023},
\href{http://arxiv.org/abs/0802.3923}{{\ttfamily arXiv:0802.3923 [hep-th]}}.

\bibitem{Zohar}
Z.~Komargodski, 2014.
\newblock {Private communication}.

\bibitem{ArkaniHamed:2001ca}
N.~Arkani-Hamed, A.~G. Cohen, and H.~Georgi, ``{(De)constructing dimensions},''
  \href{http://dx.doi.org/10.1103/PhysRevLett.86.4757}{{\em Phys.Rev.Lett.}
  {\bfseries 86} (2001) 4757--4761},
\href{http://arxiv.org/abs/hep-th/0104005}{{\ttfamily arXiv:hep-th/0104005
  [hep-th]}}.

\bibitem{Hill:2000mu}
C.~T. Hill, S.~Pokorski, and J.~Wang, ``{Gauge invariant effective Lagrangian
  for Kaluza-Klein modes},''
  \href{http://dx.doi.org/10.1103/PhysRevD.64.105005}{{\em Phys.Rev.}
  {\bfseries D64} (2001) 105005},
\href{http://arxiv.org/abs/hep-th/0104035}{{\ttfamily arXiv:hep-th/0104035
  [hep-th]}}.

\bibitem{ArkaniHamed:2001nc}
N.~Arkani-Hamed, A.~G. Cohen, and H.~Georgi, ``{Electroweak symmetry breaking
  from dimensional deconstruction},''
  \href{http://dx.doi.org/10.1016/S0370-2693(01)00741-9}{{\em Phys.Lett.}
  {\bfseries B513} (2001) 232--240},
\href{http://arxiv.org/abs/hep-ph/0105239}{{\ttfamily arXiv:hep-ph/0105239
  [hep-ph]}}.

\bibitem{Hill:2002me}
C.~T. Hill and A.~K. Leibovich, ``{Deconstructing 5-D QED},''
  \href{http://dx.doi.org/10.1103/PhysRevD.66.016006}{{\em Phys.Rev.}
  {\bfseries D66} (2002) 016006},
\href{http://arxiv.org/abs/hep-ph/0205057}{{\ttfamily arXiv:hep-ph/0205057
  [hep-ph]}}.

\end{thebibliography}\endgroup
\bibliographystyle{utphys}
\end{document}